\documentclass[twocolumn,tighten]{aastex62}

\usepackage{amsmath}

\newcommand{\dt}[1]{\textcolor{black}{#1}}
\newcommand{\fix}[1]{\textcolor{red}{#1}}
\begin{document}

\title{The Disk Substructures at High Angular Resolution Project (DSHARP):     \\ 
I. Motivation, Sample, Calibration, and Overview }

\author{Sean~M.~Andrews}
\affiliation{Harvard-Smithsonian Center for Astrophysics, 60 Garden Street, Cambridge, MA 02138, USA}

\author{Jane Huang}
\affiliation{Harvard-Smithsonian Center for Astrophysics, 60 Garden Street, Cambridge, MA 02138, USA}

\author{Laura M.~P{\'e}rez}
\affiliation{Departamento de Astronom{\'i}a, Universidad de Chile, Camino El Observatorio 1515, Las Condes, Santiago, Chile}

\author{Andrea Isella}
\affiliation{Department of Physics and Astronomy, Rice University, 6100 Main Street, Houston, TX 77005, USA}

\author{Cornelis P.~Dullemond}
\affiliation{Zentrum f{\"u}r Astronomie, Heidelberg University, Albert Ueberle Str.~2, 69120 Heidelberg, Germany}

\author{Nicol{\'a}s T.~Kurtovic}
\affiliation{Departamento de Astronom{\'i}a, Universidad de Chile, Camino El Observatorio 1515, Las Condes, Santiago, Chile}

\author{Viviana V.~Guzm{\'a}n}
\affiliation{Joint ALMA Observatory, Avenida Alonso de C{\'o}rdova 3107, Vitacura, Santiago, Chile}
\affiliation{Instituto de Astrof{\'i}sica, Pontificia Universidad Cat{\'o}lica de Chile, Av.~Vicu{\~n}a Mackenna 4860, 7820436 Macul, Santiago, Chile}

\author{John M.~Carpenter}
\affiliation{Joint ALMA Observatory, Avenida Alonso de C{\'o}rdova 3107, Vitacura, Santiago, Chile}

\author{David J.~Wilner}
\affiliation{Harvard-Smithsonian Center for Astrophysics, 60 Garden Street, Cambridge, MA 02138, USA}

\author{Shangjia Zhang}
\affiliation{Department of Physics and Astronomy, University of Nevada, Las Vegas, 4505 S.~Maryland Pkwy, Las Vegas, NV 89154, USA}

\author{Zhaohuan Zhu}
\affiliation{Department of Physics and Astronomy, University of Nevada, Las Vegas, 4505 S.~Maryland Pkwy, Las Vegas, NV 89154, USA}

\author{Tilman Birnstiel}
\affiliation{University Observatory, Faculty of Physics, Ludwig-Maximilians-Universit\"at M\"unchen, Scheinerstr.~1, 81679 Munich, Germany}

\author{Xue-Ning Bai}
\affiliation{Institute for Advanced Study and Tsinghua Center for Astrophysics, Tsinghua University, Beijing 100084, China}

\author{Myriam Benisty}
\affiliation{Unidad Mixta Internacional Franco-Chilena de Astronom\'{i}a, CNRS/INSU UMI 3386, Departamento de Astronom{\'i}a, Universidad de Chile, Camino El Observatorio 1515, Las Condes, Santiago, Chile}
\affiliation{Univ.~Grenoble Alpes, CNRS, IPAG, 38000 Grenoble, France}

\author{A.~Meredith Hughes}
\affiliation{Department of Astronomy, Van Vleck Observatory, Wesleyan University, 96 Foss Hill Drive, Middletown, CT 06459, USA}

\author{Karin I.~{\"{O}}berg}
\affiliation{Harvard-Smithsonian Center for Astrophysics, 60 Garden Street, Cambridge, MA 02138, USA}

\author{Luca Ricci}
\affiliation{Department of Physics and Astronomy, California State University Northridge, 18111 Nordhoff Street, Northridge, CA 91130, USA}

\begin{abstract}
We introduce the Disk Substructures at High Angular Resolution Project (DSHARP), one of the initial Large Programs conducted with the Atacama Large Millimeter/submillimeter Array (ALMA).  The primary goal of DSHARP is to find and characterize substructures in the spatial distributions of solid particles for a sample of 20 nearby protoplanetary disks, using very high resolution ($\sim$0\farcs035, or 5 au\dt{, FWHM}) observations of their 240 GHz (1.25 mm) continuum emission.  These data provide a first homogeneous look at the small-scale features in disks that are directly relevant to the planet formation process, quantifying their prevalence, morphologies, spatial scales, spacings, symmetry, and amplitudes, for targets with a variety of disk and stellar host properties.  We find that these substructures are ubiquitous in this sample \dt{of large, bright disks}.  They are most frequently manifested as concentric, narrow emission rings and depleted gaps, although large-scale spiral patterns and small arc-shaped azimuthal asymmetries are also present in some cases.  These substructures are found at a wide range of disk radii (from a few au to more than 100 au), are usually \dt{compact} ($\lesssim 10$ au), and show a wide range of amplitudes (brightness contrasts).  Here we discuss the motivation for the project, describe the survey design and the sample properties, detail the observations and data calibration, highlight some basic  results, and provide a general overview of the key conclusions that are presented in more detail in a series of accompanying articles.  The DSHARP data -- including visibilities, images, calibration scripts, and more -- are released for community use at \fix{\url{https://almascience.org/alma-data/lp/DSHARP}}. 
\end{abstract}

\section{Introduction and Motivation \label{sec:intro}}

There is a long-standing desire to link the properties of circumstellar disks with the initial conditions of planetary systems.  The theoretical aspiration in the field is to develop a deterministic framework that takes a set of measured disk properties \citep[e.g., the spatial distribution of densities and temperatures;][]{andrews09,andrews10b,isella09,isella10} and predicts the key characteristics of the exoplanet population \citep[e.g., masses, orbital architectures, atmosphere compositions;][]{ida04,ida08,alibert05,mordasini09}.  A quality reproduction in this population synthesis context requires the tuning of increasingly sophisticated models for the formation of planetary systems, their interactions with disk material, and their subsequent long-term dynamical evolution \citep[see][]{benz14}.    

The most crucial obstacle in the planet formation process is the assembly of planetesimals \citep[see][]{johansen14}.  The formation of terrestrial planets and giant planet cores hinges on the rapid agglomeration of small particles into these much larger ($\gtrsim$ km-sized) bodies.  Astronomers have worked on this topic and its pitfalls for more than 50 years, although without much observational guidance.  Fortunately, that is changing.  Resolved observations of the continuum emission from mm/cm-sized particles in disks measure how the solids are distributed\dt{.  Resolved variations in the continuum spectrum shape have been interpreted as} radial gradients in \dt{the particle} size distributions \citep[larger \dt{solids} closer to the star;][]{isella10,guilloteau11,perez12,lperez15,menu14,tazzari16,tripathi18}\dt{.  Pronounced discrepancies between the spatial distributions of continuum and spectral line emission have led to suggestions that the} mass ratio \dt{of solids} relative to gas \dt{also varies with radius} (higher closer to the star; \citealt{panic09,andrews12,degregorio-monsalvo13,rosenfeld13b,zhang14,facchini17,ansdell18}).  Those \dt{results} provide strong \dt{qualitative} support for evolutionary models of solids early in the planetesimal assembly process \citep[e.g.,][]{birnstiel14,testi14,birnstiel16}.  

Despite that progress, there is still considerable tension regarding planetesimal formation timescales for the default assumption of a smooth gas disk (with pressure, $P$, decreasing monotonically with radius, $r$).  This tension is associated with radial drift, the inward migration of solids toward the global $P$ maximum that occurs when they decouple from the sub-Keplerian gas flow \citep{adachi76,weidenschilling77,nakagawa86}.  The predicted drift rates for mm/cm solids located tens of au from the host star are fast enough to severely limit planetesimal growth  \citep{takeuchi02,takeuchi05,brauer07,brauer08} and are in conflict with routine observations of emission from those particles at $r \approx 10$--100 au  \citep[e.g.,][]{tripathi17,tazzari17,barenfeld17,andrews18}.  

This contradiction indicates that the $P(r)$ profiles in disks are likely not smooth.  Localized $P$ modulations can slow or trap drifting solids \citep{whipple72,pinilla12b}, perhaps concentrating them enough to trigger gravitational and/or streaming instabilities that rapidly convert pebbles to planetesimals \citep[e.g.,][]{youdin02,youdin05,johansen09}.  Such particle traps or other migration bottlenecks could be produced by the dynamics associated with how gas, dust, and magnetic fields are coupled \citep[e.g.,][]{dzyurkevich13,bai14b,flock15,lyra15,dipierro15,bethune17,dullemond18,suriano18} or by strong gradients in material properties (e.g., \citealt{okuzumi12,estrada16,armitage16,stammler17,pinilla17}\dt{; but see \citealt{vanterwisga18,long18}}).  These small-scale material concentrations -- {\it substructures} -- are largely absent in contemporary models of planet formation, but they would likely play fundamental roles in nearly all aspects of the formation process.

If such substructures were prominent in disks at early evolution stages, it is possible that planetesimals and even entire planetary systems were created much more efficiently than is expected in the traditional models \citep[e.g.,][]{greaves10,najita14,nixon18}.  In that scenario, the typical $\sim$Myr-old disk may harbor a `second generation' of substructures created by the dynamical interactions between young planets and their nascent disk material \citep[see][]{lin93,kley12}, which in turn can affect the orbital architectures of those burgeoning planetary systems \citep[e.g.,][]{coleman16}.  

In any case, observations of disk substructures are essential.  Direct constraints on small-scale gas pressure variations in disks based on high resolution measurements of molecular line emission are a formidable challenge.  However, the particle trapping capabilities of even modest pressure maxima should substantially amplify the associated local mm/cm-sized particle density \citep[e.g.,][]{paardekooper06,rice06,pinilla12a,zhu12}, generating a bright signature in the broadband (sub-)mm continuum that is much easier to measure on the smallest scales.  

The initial foray into such work came from the ``transition" disks \citep{strom89,skrutskie90,calvet02}, which show dense particle rings at $r \approx$~tens of au, outside depleted central cavities \citep[e.g.,][]{andrews11,vandermarel18,pinilla18}.  Observations with sufficient resolution reveal that these particle traps exhibit complex substructures, including azimuthal asymmetries \citep{casassus13,vandermarel13,isella13,perez14}, additional rings \citep{fedele17,vanderplas17a}, warped geometries \citep[and/or radial inflows;][]{rosenfeld12,rosenfeld14,marino15,casassus18}, and spiral arms \citep{christiaens14,boehler18,dong18}.  Similar features have been identified from the IR starlight scattered off the disk atmospheres \citep[e.g.,][]{muto12,grady13,quanz13,avenhaus14,rapson15,benisty15,deboer16,ginski16,akiyama16}.  

Some serendipitous discoveries at modest ($\sim$10--20 au) resolution hint that the more general disk population frequently exhibits substructures in the forms of rings/gaps \citep{zhang16,isella16,cieza16,cieza17,loomis17,huang17,cox17,dipierro18,fedele18,vanterwisga18} and spirals \citep{perez16}.  The richness of these substructures becomes clear for the few individual cases that have had their continuum emission probed at resolutions of only a few au (HL Tau, \citealt{brogan15}; TW Hya, \citealt{andrews16}; MWC 758, \citealt{dong18}).  Again, similar conclusions are being drawn from complementary measurements of scattered light from small dust grains \citep[e.g.,][]{vanboekel17,avenhaus18}.

All of these observations suggest that substructures are common, and therefore are likely significant factors in many disk evolution and planet formation processes.  Moreover, they demonstrate a tremendous opportunity: high resolution mm continuum measurements can quantify the forms, prevalence, and diversity (e.g., in scales, locations, amplitudes) of disk substructures, and thereby help develop a more robust theoretical framework for characterizing the early evolution of planetary systems.  The next step along that path is to move from a serendipitous discovery-space to a principled survey specifically designed to study these features.        

In this article, we introduce a new survey that moves in this direction.  The Disk Substructures at High Angular Resolution Project (DSHARP) was conducted as one of the first ALMA Large Programs.  DSHARP measures the 240 GHz continuum emission at $\sim$35 mas (5 au) resolution for 20 disks, to help better understand the evolution of solid particles during the planet formation process.  Having motivated the project, this article also describes the DSHARP survey design and sample (Section~\ref{sec:sample}), the ALMA observations (Section~\ref{sec:obs}) and their calibration (Section~\ref{sec:cal}), along with some basic observational results and the DSHARP data release (Section~\ref{sec:results}).  We conclude with an overview of the highlights from a series of accompanying articles (Section~\ref{sec:discussion}).

\section{Survey Design and Sample \label{sec:sample}}

\begin{deluxetable*}{l l c c c c c c c c c}
\tabletypesize{\scriptsize}
\tablecaption{DSHARP Sample: Host Star Properties \label{table:stars}}
\tablehead{
\colhead{Name} & \colhead{Region} & 2MASS & \colhead{$d$} & \colhead{SpT} & 
\colhead{$\log{T_{\rm eff}}$} & \colhead{$\log{L_\ast}$} & \colhead{$\log{M_\ast}$} & \colhead{$\log{t_\ast}$} & \colhead{$\log{\dot{M}_\ast}$} & \colhead{Refs.}  
\\
\colhead{} & \colhead{} & \colhead{designation} & \colhead{(pc)} & \colhead{} & 
\colhead{(K)} & \colhead{($L_\odot$)} & \colhead{($M_\odot$)} & 
\colhead{(yr)} & \colhead{$M_\odot \, {\rm yr}^{-1}$} & \colhead{}
}
\colnumbers
\startdata
HT Lup \tablenotemark{a} & Lup I      & J15451286-3417305 & $154\pm2$ & K2   & $3.69\pm0.02$ 
    & \phd$0.74\pm0.20$ & \phd0.23 $^{+0.06}_{-0.13}$ & $5.9\pm0.3$          & $<$ -8.4     & 1, 1, 1     \\
GW Lup                   & Lup I      & J15464473-3430354 & $155\pm3$ & M1.5 & $3.56\pm0.02$ 
    & -$0.48\pm0.20$    & -0.34 $^{+0.10}_{-0.17}$    & $6.3\pm0.4$          & -$9.0\pm0.4$ & 1, 1, 1     \\
IM Lup                   & Lup II     & J15560921-3756057 & $158\pm3$ & K5   & $3.63\pm0.03$ 
    & \phd$0.41\pm0.20$ & -0.05 $^{+0.09}_{-0.13}$    & $5.7\pm0.4$          & -$7.9\pm0.4$ & 1, 1, 1     \\
RU Lup                   & Lup II     & J15564230-3749154 & $159\pm3$ & K7   & $3.61\pm0.02$ 
    & \phd$0.16\pm0.20$  & -0.20 $^{+0.12}_{-0.11}$    & $5.7\pm0.4$          & -$7.1\pm0.3$ & 1, 1, 1    \\
Sz 114                   & Lup III    & J16090185-3905124 & $162\pm3$ & M5   & $3.50\pm0.01$ 
    & -$0.69\pm0.20$     & -0.76 $^{+0.08}_{-0.07}$    & 6.0 $^{+0.1}_{-0.8}$ & -$9.1\pm0.3$ & 1, 1, 1    \\
Sz 129                   & Lup IV     & J15591647-4157102 & $161\pm3$ & K7   & $3.61\pm0.02$ 
    & -$0.36\pm0.20$     & -0.08 $^{+0.03}_{-0.15}$    & $6.6\pm0.4$          & -$8.3\pm0.3$ & 1, 1, 1    \\
MY Lup \tablenotemark{b} & Lup IV     & J16004452-4155310 & $156\pm3$ & K0   & $3.71\pm0.02$ 
    & -$0.06\pm0.20$     & \phd0.09 $^{+0.03}_{-0.13}$ & 7.0 $^{+0.6}_{-0.3}$ & $<$ -9.6     & 1, 1, 1    \\
HD 142666                & Upper Sco  & J15564002-2201400 & $148\pm2$ & A8   & $3.88\pm0.02$ 
    & \phd$0.96\pm0.21$  & \phd0.20 $^{+0.04}_{-0.01}$ & $7.1\pm0.3$          & $<$ -8.4     & 2, 2, 2    \\
HD 143006                & Upper Sco  & J15583692-2257153 & $165\pm5$ & G7   & $3.75\pm0.02$ 
    & \phd$0.58\pm0.15$  & \phd0.25 $^{+0.05}_{-0.08}$ & $6.6\pm0.3$          & -$8.1\pm0.4$ & 3, 4, 5    \\
AS 205 \tablenotemark{a} & Upper Sco  & J16113134-1838259 & $128\pm2$ & K5   & $3.63\pm0.03$ 
    & \phd$0.33\pm0.15$  & -0.06 $^{+0.07}_{-0.05}$    & $5.8\pm0.3$          & -$7.4\pm0.4$ & 3, 4, 6    \\
SR 4                     & Oph L1688  & J16255615-2420481 & $134\pm2$ & K7   & $3.61\pm0.02$ 
    & \phd$0.07\pm0.20$  & -0.17 $^{+0.11}_{-0.14}$    & $5.9\pm0.4$          & -$6.9\pm0.5$ & 7, 8, 9    \\
Elias 20                 & Oph L1688  & J16261886-2428196 & $138\pm5$ & M0   & $3.59\pm0.03$ 
    & \phd$0.35\pm0.20$  & -0.32 $^{+0.12}_{-0.07}$    & $<5.9$               & -$6.9\pm0.5$ & 9, 10, 9   \\
DoAr 25                  & Oph L1688  & J16262367-2443138 & $138\pm3$ & K5   & $3.63\pm0.03$ 
    & -$0.02\pm0.20$     & -0.02 $^{+0.04}_{-0.19}$    & $6.3\pm0.4$          & -$8.3\pm0.5$ & 11, 10, 12 \\
Elias 24                 & Oph L1688  & J16262407-2416134 & $136\pm3$ & K5   & $3.63\pm0.03$ 
    & \phd$0.78\pm0.20$  & -0.11 $^{+0.16}_{-0.08}$    & $5.3\pm0.4$          & -$6.4\pm0.5$ & 11, 8, 9   \\
Elias 27                 & Oph L1688  & J16264502-2423077 & 116 $^{+19}_{-10}$ & M0  & $3.59\pm0.03$ 
    & -$0.04\pm0.23$     & -0.31 $^{+0.15}_{-0.11}$    & $5.9\pm0.5$          & -$7.2\pm0.5$ & 7, 10, 9   \\
DoAr 33                  & Oph L1688  & J16273901-2358187 & $139\pm2$ & K4   & $3.65\pm0.03$ 
    & \phd$0.18\pm0.20$  & \phd0.04 $^{+0.05}_{-0.17}$ & $6.2\pm0.4$          & \nodata & 13, 8           \\
WSB 52                   & Oph L1688  & J16273942-2439155 & $136\pm3$ & M1   & $3.57\pm0.03$ 
    & -$0.15\pm0.20$     & -0.32 $^{+0.13}_{-0.17}$    & $5.8\pm0.5$          & -$7.6\pm0.5$ & 7, 8, 9    \\
WaOph 6                  & Oph N 3a   & J16484562-1416359 & $123\pm2$ & K6   & $3.62\pm0.03$ 
    & \phd$0.46\pm0.20$  & -0.17 $^{+0.17}_{-0.09}$    & $5.5\pm0.5$          & -$6.6\pm0.5$ & 14, 10, 14 \\
AS 209                   & Oph N 3a   & J16491530-1422087 & $121\pm2$ & K5   & $3.63\pm0.03$ 
    & \phd$0.15\pm0.20$  & -0.08 $^{+0.11}_{-0.14}$    & $6.0\pm0.4$          & -$7.3\pm0.5$ & 15, 10, 6  \\
HD 163296                & isolated?  & J17562128-2157218 & $101\pm2$ & A1   & $3.97\pm0.03$ 
    & \phd$1.23\pm0.30$  & \phd0.31 $^{+0.05}_{-0.03}$ & $7.1\pm0.6$          & -$7.4\pm0.3$ & 2, 2, 2    \\
\enddata
\tablecomments{Col.~(1) Target name. Col.~(2) Associated star-forming region.  The Lup sub-cloud regions are as designated by \citet{cambresy99}.  Upper Sco memberships were made following \citet{luhman18}.  AS 209 and WaOph 6 are located well northeast of the main Oph region in the Oph N 3a complex.  They are most closely associated with the L163 and L162 dark clouds, respectively.  Col.~(3) The 2MASS designations, to aid in catalog cross-referencing.  Col.~(4) Distance (computed from the {\it Gaia} DR2 parallaxes).  Col.~(5) Spectral type from the literature (first reference entries in Col.~11).  Col.~(6) Effective temperatures from the literature (second reference entries in Col.~11).  Col.~(7) Stellar luminosities from the literature, scaled according to the appropriate $d$ in Col.~(4) (second reference entries in Col.~(11).  Cols.~(8)+(9) Stellar masses and ages.  Col.~(9) Accretion rates, inferred from (properly scaled) accretion luminosities (third reference entries in Col.~11).  All quoted measurements correspond to the peak of the marginalized posterior distributions.  Uncertainties reflect the 68.3\%\ confidence interval; limits are taken at the 95.5\%\ confidence level.}
\tablenotetext{a}{HT Lup (Sz 68) and AS 205 (V866 Sco) are triple systems.  See \citet{troncoso18} for details.}
\tablenotetext{b}{The MY Lup disk is inclined and flared enough that it likely extincts the host: the $L_\ast$ and $t_\ast$ estimates may be too faint and old, respectively.}
\tablerefs{In Col.~(11), the references for the quoted SpT, \{$T_{\rm eff}$, $L_\ast$\}, and accretion luminosity measurements, respectively: 1 = \citet{alcala17}, 2 = \citet{fairlamb15}, 3 = \citet{luhman12}, 4 = \citet{barenfeld16}, 5 = \citet{rigliaco15}, 6 = \citet{salyk13}, 7 = \citet{luhman99b}, 8 = \citet{andrews10b}, 9 = \citet{natta06}, 10 = \citet{andrews09}, 11 = \citet{wilking05}, 12 = \citet{muzerolle98}, 13 = \citet{bouvier92}, 14 = \citet{eisner05}, 15 = \citet{hbc88}.}
\end{deluxetable*}

The DSHARP survey was designed to optimize the spatial resolution and contrast sensitivity to continuum emission substructures.  Secondarily, measurements of CO line emission were also of interest as a preliminary opportunity to identify corresponding gas structures and infer other relevant bulk disk properties (e.g., geometry).  We defined two criteria to guide the survey design, based on previous observations and theoretical expectations for the origins of disk substructures.

The first criterion was access to a wide range of spatial scales down to a \dt{FWHM} resolution of $\sim$5 au.  Such high resolution was essential for identifying the disk substructures in the sharpest ALMA continuum images available to date \citep{brogan15,andrews16}.  Moreover, it is comparable to the (disk-averaged) pressure scale height, $h_P$ (where $h_P/r \approx 0.1$; \citealt{kenyon87}), a benchmark size that is directly related to the $P$ deviations generated by turbulent zonal flows \citep[e.g.,][]{johansen09}, vortices \citep[e.g.,][]{barge95}, or planetary gaps \citep[e.g.,][]{bryden99}.  At 5 au resolution, $h_P$-sized features in radius or azimuth are resolved in the outer disk, and detectable down to $r \approx 10$ au (for sufficient contrast).    

The second criterion was the ability to detect a $\sim$10\%\ contrast out to Solar System size-scales ($r \approx 40$ au).  This is roughly the contrast measured for the weaker substructures in the HL Tau and TW Hya disks \citep[e.g.,][]{akiyama16,huang18}.  It is also sufficient to detect the continuum emission that (indirectly) traces the $\sim$20\%\ pressure variations produced by $\gtrsim 0.1$ $M_{\rm Jup}$ planets \citep{fung14}, zonal flows \citep{simon14}, or weak vortices \citep[e.g.,][]{goodman87}, even if (contrary to expectations) there is no accompanying amplification in the concentration of the solids (presuming the emission is optically thin). 

The combination of these criteria and ALMA technical restrictions meant that the optimal observing frequency was in the vicinity of 240 GHz (Band 6).  Higher frequency observations at comparable (or better) resolution were not permitted for Cycle 4 Large Programs, and the resolution and sensitivity options at lower frequencies were both insufficient for our goals.   

The resolution criterion drove planning for the survey sample.  The Cycle 4 configuration schedule was set to provide the requisite resolution (with baseline lengths out to 6.8--12.6 km) during 2017 June and July.  We targeted disks that are nearby enough to give the required spatial resolution for those configurations, and that transit at high elevations at night during this period.  This limited the sample pool to the Oph \citep{wilking08}, Lup \citep{comeron08}, and Upper Sco \citep{preibisch08} regions, plus a few isolated targets.  The field was narrowed to focus on Class II sources to avoid confusion with envelope emission.  We excluded ``transition" disks, since they are already known to exhibit substructures (by definition).

Those criteria leave $\sim$200 viable targets.  A more severe cut was then made to meet the contrast criterion.  A general framing of that criterion is somewhat arbitrary, but we chose some fiducial numbers as a guide.  For a target at 140 pc and with a synthesized beam FWHM of 35 mas, we aimed to measure a 10\%\ deviation from an otherwise smooth brightness profile (at SNR $\ge$ 2 per beam) out at $r = 40$ au ($\sim$0\farcs3).  This metric requires previous continuum observations at modest (0\farcs3) resolution for selection \citep{andrews09,andrews10b,ansdell16,barenfeld16}.  For reasonable assumptions about the shape of the brightness profile,\footnote{We conservatively assumed a face-on orientation with $I_\nu \propto r^{-0.5}$--$r^{-1}$ \citep[see][]{tripathi17,andrews18}.} this criterion can be met with a cut on the 0\farcs3 peak brightness.  Experimentation with simulated data suggested a peak brightness cut at 20 mJy per 0\farcs3 beam (4.8 K) is appropriate, implying an objective noise level of 17 $\mu$Jy per 35 mas beam (0.3 K).\footnote{Unless specified otherwise, DSHARP brightness temperatures are calculated assuming the standard Rayleigh-Jeans relation.}  The caveat is that much of the available data at 0\farcs3 resolution were taken at 340 GHz; in the applicable cases, we assumed that $I_\nu \propto \nu^{2.5}$ \citep[cf.,][]{aw05}.       

While that brightness cut substantially reduces the pool, the sample size was ultimately set by ALMA restrictions.  Only $\sim$30 hours in the LST ranges of interest were set aside for Large Programs in each of the two relevant array configurations.  The desired noise could be reached in $\sim$1 hour of integration per target, but the factor of three overhead costs meant that the sample size was limited to 10 targets per configuration.  We selected 10 targets (mostly) in Oph for the more compact of the two configurations (C40-8, $\le 6.8$ km baselines; 50 mas resolution), based on their nominally closer distances \citep[125 pc;][]{degeus89,loinard08}.\footnote{Note that the 125 pc distance used to motivate the slightly coarser resolution for Oph targets was inappropriate.  However, this aspect of the survey design was ignored anyway, due to unforeseen alterations in the configuration schedule.}  Ten more targets (primarily in Lup) were chosen for C40-9 ($\le 12.6$ km baselines; 35 mas resolution).        

\begin{figure*}[t!]
\includegraphics[width=\linewidth]{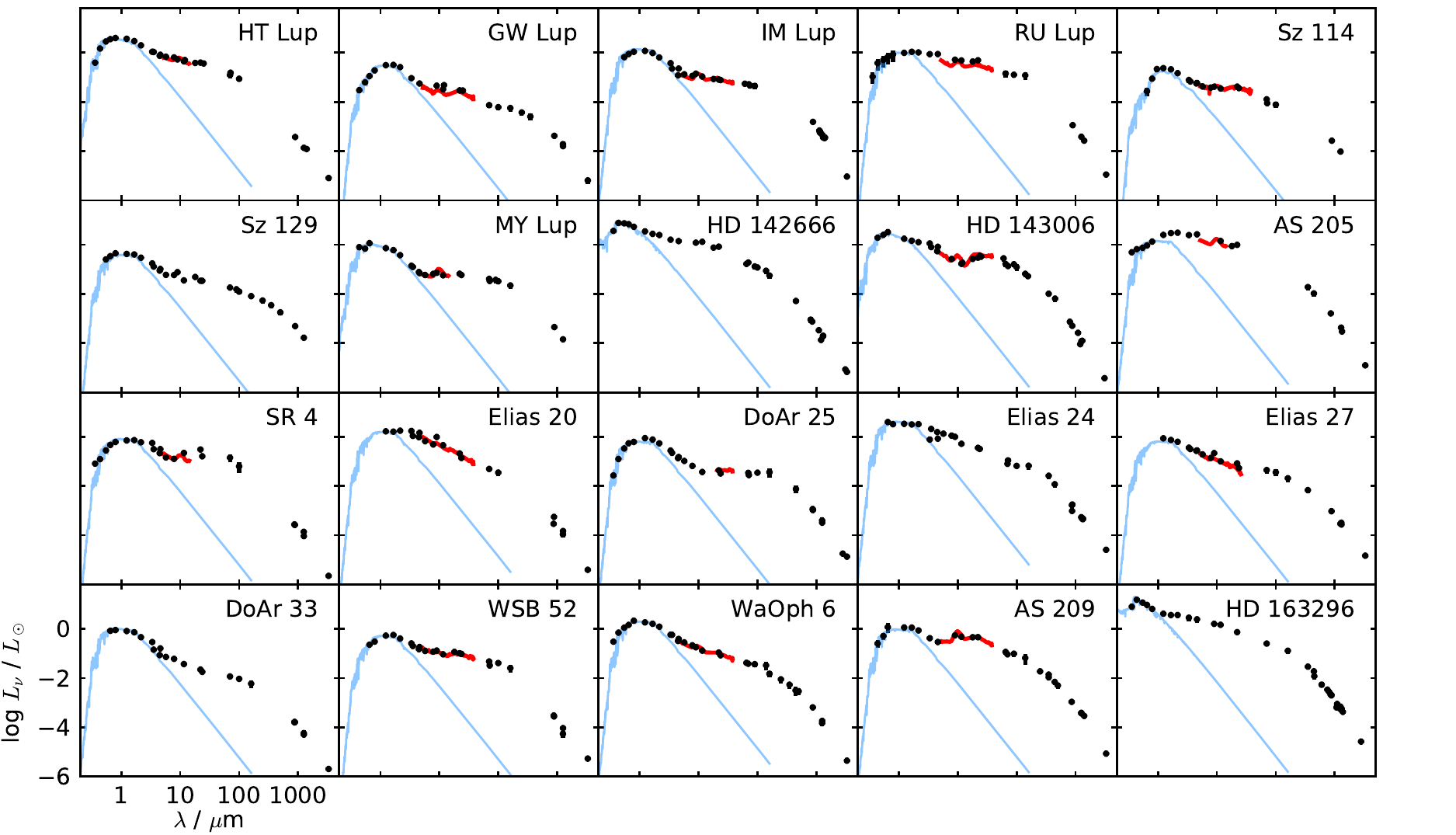}
\figcaption{Broadband SEDs for the DSHARP targets.  The ordinate is $L_\nu = 4 \pi d^2 \nu F_\nu$ in $L_\odot$ units.  These SEDs have been de-reddened using the extinction values quoted by the references in Col.~(11) of Table~\ref{table:stars} (second entries) and the prescription described by \citet{andrews13}.  Blue curves show the {\sc Nextgen}/{\tt BT-settl} photosphere models \citep{allard03,allard11} corresponding to the stellar parameters listed in Table~\ref{table:stars}.  Red curves show the {\it Spitzer} IRS spectra.  Note that the SEDs for HT Lup and AS 205 include contributions from multiple components.  Optical photometry was collected from a range of sources \citep{vrba93,herbst94,hughes94,oudmaijer01,wilking05,gras-velazquez05,eisner05,padgett06,grankin07,merin08,mendigutia12}; infrared data were culled from 2MASS \citep{skrutskie06}, {\it WISE} \citep{wright10}, {\it Spitzer} imaging surveys \citep{carpenter08,evans09}, {\it AKARI} \citep{ishihari10}, and {\it Herschel} (IRSA); (sub-)mm data come from various sources \citep{andre94,mannings94,sylvester96,nurnberger97,mannings97,dent98,henning98,natta04,stanke06,aw07b,lommen07,lommen09,roccatagliata09,andrews09,isella07,pinte08,isella09,ricci10b,sandell11,oberg11,perez12,lperez15,qi15,ansdell16,cleeves16,barenfeld16,ubach17,huang17,tripathi17,cox17,andrews18}.  These SEDs are available in the DSHARP data release.         
\label{fig:SEDs}}
\end{figure*}

The resulting sample \dt{and its} stellar host properties are compiled in Table~\ref{table:stars}.  Target distances ($d$) were derived from {\it Gaia} DR2 parallax measurements \citep{gaia_dr2}, following \citet{astraatmadja16} for a flat $d$ prior.  Literature estimates of the effective temperatures ($T_{\rm eff}$) and luminosities ($L_\ast$; re-scaled for the appropriate $d$) were adopted to derive masses ($M_\ast$) and ages ($t_\ast$) based on the {\tt MIST} models \citep{choi16}, following the methodology described by \citet{andrews18}.  Accretion rates ($\dot{M}_\ast$) were calculated from those host parameters and literature measurements of accretion luminosities (scaled for $d$; see Table~\ref{table:stars}).  The sample hosts exhibit \dt{a} range of young star properties, with $M_\ast \approx 0.2$--2 $M_\odot$ and nearly two decades spanned in both $L_\ast$ and $\dot{M}_\ast$.  The mean age is 1 Myr, although with considerable individual uncertainties (and various untreated systematics; see \citealt{soderblom14}).  

The broadband spectral energy distributions (SEDs) for the sample are shown together in Figure~\ref{fig:SEDs}.  Relative to the median SED (normalized at 1.5 $\mu$m) of larger samples of Class II targets \citep[e.g.,][]{ribas17}, RU Lup, AS 205, and AS 209 are in the top quartile (i.e., are over-luminous); the SEDs for HD 143006, SR 4, and DoAr 25 are relatively low in the near-infrared and high in the far-infrared (similar to, though not nearly as pronounced as, the typical transition disk SED); and the SEDs for DoAr 33 and WaOph 6 are in the bottom quartile.  \dt{This diversity in the SEDs is one potential basis for future explorations of how the resolved emission distributions vary with relevant ``bulk" parameters (e.g., the amount of dust settling toward the disk midplane).}   

\dt{While these sample targets do cover a range of properties, it is worth emphasizing that this range is not representative of the general population.}  The sample hosts tend to have earlier spectral types, and are accordingly more massive, luminous, and accreting more vigorously than stars at the peak of the initial mass function.  This host bias enters implicitly with the sensitivity criterion, since we required a previous resolved measurement.  The studies that provided those data are biased toward brighter continuum sources, which permeates to the host properties since the continuum luminosity scales steeply with $M_\ast$ \citep{andrews13,mohanty13} and $\dot{M}_\ast$ \citep{manara16,mulders17}.  The same is true for multiple star systems: these were not explicitly excluded, but the selection criteria bias against them \dt{because close companions tend to reduce the system continuum emission \citep[e.g.,][]{jensen94,harris12}}.  

\dt{The bias in favor of targets with brighter continuum emission also translates into a preferential selection of {\it larger} disks, given the observed size-luminosity correlation \citep{tripathi17,tazzari17,andrews18}.  This corresponding size bias is decidedly beneficial for achieving the DSHARP goals discussed in Section~\ref{sec:intro}.  As we noted above, the general theoretical predictions for substructure sizes are comparable to the gas pressure scale height ($h_P$), which increases roughly linearly with disk radius.  For a fixed resolution, it should be easier to identify and characterize the larger substructures expected at larger disk radii.}  

\dt{To roughly quantify these biases, we can make a comparison between targets that are more representative of the general disk population and the average member of the DSHARP sample.  A ``typical" target has a host star mass near the peak of the mass function ($M_\ast \approx 0.3$ $M_\odot$, or spectral type M3--M4) and continuum emission from its disk that is both relatively faint ($F_\nu \approx 10$--15 mJy; \citealt{ansdell16,cieza19}) and compact ($R_{\rm eff} \approx 10$--20 au, with the effective radius defined by \citealt{tripathi17}; see also \citealt{andrews18}).  The DSHARP averages are $M_\ast \approx 0.8$ $M_\odot$ (spectral type K7), $F_\nu \approx 150$ mJy, and $R_{\rm eff} \approx 50$ au; only the sample extremes stretch down toward ``typical" values.}

\dt{These biases are difficult to mitigate for studies focused on finding and characterizing disk substructures, presuming their size scales are usually comparable to $h_P$.  The ``typical" disk is compact enough that $h_P$ for the radii where there is still continuum emission is smaller than the best resolutions available with ALMA.  If this is the case, then we could be left probing only the extreme large end of substructures in ``typical" disks, making any assessments of prevalence difficult (i.e., failed searches for substructures would still permit plenty of $h_P$-sized features to be present on sub-resolution scales).  One option is to push to higher frequencies and thereby better resolution, but then high optical depths would limit the discovery-space to substructures in the form of dramatic depletions (e.g., very deep gaps) only.}

\begin{deluxetable*}{lclccccc}
\tabletypesize{\scriptsize}
\tablecaption{DSHARP Observing Log (ALMA Program 2016.1.00484.L) \label{table:obs}}
\tablehead{
\colhead{Name} & 
\colhead{UTC Date} & 
\colhead{Config.} & 
\colhead{Baselines} & 
\colhead{$N_{\rm ant}$} & 
\colhead{$\mathcal{E} / \degr$} & 
\colhead{PWV/mm} & 
\colhead{Calibrators}
}
\colnumbers
\startdata
HT Lup    & 2017/05/14--04:11 & C40-5   & \phn15\,m -- \phn1.1\,km & 43 & 76--77 & 1.00--1.15 
          & J1517-2422, J1427-4206, J1610-3958, J1540-3906 \\
          & 2017/05/17--02:12 & C40-5   & \phn15\,m -- \phn1.1\,km & 49 & 58--67 & 0.90--1.05
          & J1517-2422, J1517-2422, J1610-3958, J1540-3906 \\
          & 2017/09/24--17:39 & C40-8/9 & \phn41\,m -- 12.1\,km    & 39 & 59--70 & 0.60--1.15
          & J1517-2422, J1517-2422, J1534-3526, J1536-3151 \\
          & 2017/09/24--19:12 & C40-8/9 & \phn41\,m -- 12.1\,km    & 39 & 75--78 & 0.65--1.05 
          & J1517-2422, J1427-4206, J1534-3526, J1536-3151 \\
GW Lup    & 2017/05/14--04:11 & C40-5   & \phn15\,m -- \phn1.1\,km & 43 & 69--72 & 1.00--1.15 
          & J1517-2422, J1427-4206, J1610-3958, J1540-3906 \\
\enddata
\tablecomments{Col.~(1) Target name.  Col.~(2) UTC date and time at the start of the observations.  Col.~(3) ALMA configuration.  Col.~(4) Minimum and maximum baseline lengths.  Col.~(5) Number of antennas available.  Col.~(6) Target elevation range.  Col.~(7) Range of precipitable water vapor levels.  Col.~(8) From left to right, the quasars observed for calibrating the bandpass, amplitude scale, phase variations, and checking the phase transfer.  Additional archival observations used in our analysis are compiled in Table~\ref{table:obs_archive}.  Table~\ref{table:obs} is published in its entirety in the electronic edition of the journal.  A portion is shown here for guidance regarding its form and content.}
\end{deluxetable*}

\begin{deluxetable*}{lclccccc}
\tabletypesize{\scriptsize}
\tablecaption{Archival ALMA Datasets Used by DSHARP \label{table:obs_archive}}
\tablehead{
\colhead{Name} & 
\colhead{UTC Date} & 
\colhead{Config.} & 
\colhead{Baselines} & 
\colhead{$N_{\rm ant}$} & 
\colhead{Calibrators} & 
\colhead{Program} & 
\colhead{References}
}
\colnumbers
\startdata
IM Lup    & 2014/07/06--22:18 & C34-4   & 20 -- \phn650 m & 31 & J1427-4206, Titan, J1534-3526, J1626-2951 & 2013.1.00226.S & 1 \\
          & 2014/07/17--01:38 & C34-4   & 20 -- \phn650 m & 32 & J1427-4206, Titan, J1534-3526, \nodata\phn\phn\phn\phn & 2013.1.00226.S & 1 \\
          & 2015/01/29--09:48 & C34-2/1 & 15 -- \phn349 m & 40 & J1517-2422, Titan, J1610-3958, \nodata\phn\phn\phn\phn & 2013.1.00694.S & 2 \\
          & 2015/05/13--08:30 & C34-3/4 & 21 -- \phn558 m & 36 & J1517-2422, Titan, J1610-3958, \nodata\phn\phn\phn\phn & 2013.1.00694.S & 2 \\
          & 2015/06/09--23:42 & C34-5   & 21 -- \phn784 m & 37 & J1517-2422, Titan, J1610-3958, J1614-3543              & 2013.1.00798.S & 3 \\
\enddata
\tablecomments{Col.~(1) Target name.  Col.~(2) UTC date and time at the start of the observations.  Col.~(3) ALMA configuration.  Col.~(4) Range of baseline lengths.  Col.~(5) Number of antennas available.  Col.~(6) From left to right, the quasars observed for calibrating the bandpass, amplitude scale, phase variations, and checking the phase transfer.  \dt{An entry of `$\ldots$' indicates no calibrator was observed for checking the phase transfer.}  Col.~(7) ALMA program ID.  Col.~(8) Original references for these datasets.  Table~\ref{table:obs_archive} is published in its entirety in the electronic edition of the journal.  A portion is shown here for guidance regarding its form and content.}
\tablerefs{1 = \citet{oberg15}, 2 = \citet{cleeves17}, 3 = \citet{pinte18}, 4 = \citet{salyk14}, 5 = \citet{dipierro18}, 6 = \citet{perez16}, 7 = \citet{huang16}, 8 = \citet{fedele18}, 9 = \citet{flaherty15}, 10 = \citet{isella16}.}
\end{deluxetable*}

\section{Observations \label{sec:obs}}

The DSHARP ALMA observations were conducted in 2017 May--November as part of program 2016.1.00484.L.  All measurements used the Band 6 receivers and correlated data from four spectral windows (SPWs) in dual polarization mode.  The continuum was sampled in three SPWs, centered at 232.6, 245.0, and 246.9 GHz, each with 128 channels spanning 1.875 GHz (31.25 MHz per channel).  The remaining SPW was centered at the $^{12}$CO $J$=2$-$1 rest frequency (230.538 GHz) and covered a bandwidth of 938 MHz in 3840 channels (488 kHz channel spacing, 0.64 km s$^{-1}$ velocity resolution).  The plan was to observe each target briefly in the C40-5 (hereafter ``compact") configuration, and also for $\sim$1 hour in the C40-8 or C40-9 (hereafter ``extended") configurations.  \dt{The compact observations are necessary to recover emission on the larger angular scales that are not sampled in the extended configurations.}  The actual observing log is provided in Table~\ref{table:obs}.   

The compact observations used an array with baseline lengths from 15 m to 1.1 km (a resolution of $\sim$0\farcs25).  The FWHM continuum and CO (per channel) emission scales are $\lesssim$ 2\arcsec, so spatial filtering should be negligible \citep[cf.,][]{wilner94}.  These observations cycled between nearby targets and totaled $\sim$12 minutes of integration time per source.  A nearby phase calibrator was observed every 6 minutes; an additional ``check" calibrator (to assess the quality of phase transfer) was observed every 30 minutes.  A bandpass and amplitude calibrator (sometimes the same quasar) were observed during each observing block.  The log in Table~\ref{table:obs} includes information about the observing conditions and calibrators.

We relied on archival ALMA observations of 5 targets (IM Lup, HD 142666, Elias 24, Elias 27, HD 163296) instead of obtaining new compact data, and folded in archival data for 3 other targets (HD 143006, AS 205, AS 209).  Information about these datasets are compiled in Table~\ref{table:obs_archive}.  The setups, observing strategies, and weather conditions are described in the listed references.

Due to a long stretch of inclement weather, the extended configuration observations were delayed until 2017 September, and continued through November.  Despite the non-optimal scheduling for the DSHARP sample, nearly all of the targets were observed for two executions (often in different configurations) in good conditions, each with $\sim$35 minutes of on-source integration time.  Sz 114, AS 205, and DoAr 25 each had only a single successful execution.  The spectral setup was the same as for the compact datasets.  Observations cycled between a single target and a nearby phase calibrator on 1 minute intervals, with a ``check" calibrator visited every 30 minutes.  Bandpass and amplitude calibrators were observed in each execution block (see Table~\ref{table:obs}).

\section{Calibration \dt{and Imaging} \label{sec:cal}}

Since the DSHARP survey is among the first to collect a large volume of ALMA data on such long baselines, a substantial effort was made to explore various calibration strategies to enhance the data quality.  The standard methodology we adopted is described here.  The specific details on the calibration of datasets for individual targets (i.e., calibration scripts) are available in the DSHARP data release (see Section~\ref{sec:results}).  All calibration tasks are performed with the {\tt CASA} package \citep{mcmullin07} and a small supplement of {\tt python} tasks.

\subsection{Pipeline Calibration}

The first step was a standard ALMA pipeline calibration.  This procedure was performed by ALMA staff separately for the compact and extended data, using {\tt CASA} {\tt v4.7.2} or {\tt v5.1.1} for datasets that were processed before or after 2017 November, respectively.  The pipeline imports the raw data and flags problematic scans, channels, or antennas.  It then derives a table of system temperatures ($T_{\rm sys}$).  Most of the DSHARP data have $T_{\rm sys} \approx 60$--80 K; in the poorest conditions it reached 130 K, and in the best cases it was 50 K.  Next, the pipeline adjusts the visibility phases according to water vapor radiometer (WVR) measurements.  For the extended data, the WVR corrections improved the median RMS phase variations by a factor of $\sim$1.7, although individual datasets saw improvements between 1.2--3.  The corrected RMS phase variations (far from the reference antenna) were typically 30\degr\ (with a range $\sim$15--50\degr).  The compact observations saw similar improvement factors (1.5--3.0) and RMS phase variations ($\sim$10\degr).                  

The pipeline then performs a bandpass calibration, using the first quasar in the calibrator list in Table~\ref{table:obs}.  It continues by setting the amplitude scale, using measurements of the second quasar in the Table~\ref{table:obs} list.  The flux density in each SPW for that quasar is determined from a power-law spectral model based on bi-monthly monitoring in ALMA Bands 3 and 7 ($\sim$100 and 340\,GHz) that is tied to primary calibrators (planets or moons).  Finally, the gain variations with time are corrected, with reference to repeated measurements of the nearby quasar listed third in the Table~\ref{table:obs} calibrator list.

\subsection{Self-Calibration} \label{sec:selfcal}

We next performed some substantial post-processing, with particular emphasis on combining datasets (from different array configurations and observations) and self-calibrating the visibilities.  We generally followed the homogenized strategy described below, using {\tt CASA} {\tt v5.1.1} and a set of custom {\tt python} routines.

The procedure started with the compact data.  A pseudo-continuum dataset was created by flagging data within $\pm$25 km s$^{-1}$ from the CO $J$=2$-$1 line center and averaging into 125 MHz channels.  The visibilities corresponding to each individual observation were imaged (Section~\ref{sec:imaging}) and checked to ensure consistent astrometric registration and flux calibration (if necessary, they are corrected; see Section~\ref{sec:align}).  The individual datasets were then re-combined.  Next, we performed a series of phase-only self-calibration iterations, stepping down the solution interval (60, 30, 18, and 6 s).  Reference antennas were selected based on data quality and proximity to the array center.  When possible, we avoided combining SPWs (or scans) to correct for SPW-dependent gain variations.  After each iteration, the data were imaged.  A noise estimate was made in an annular region within a 4\farcs25-radius circle centered on the target but excluding the image mask.  This self-calibration sequence is stopped after reaching a solution interval on the record length (6 s) or if the peak SNR does not increase by $> 5$\%\ from the previous iteration.  Finally, we performed one iteration of amplitude self-calibration (for each SPW independently) on a scan interval ($\sim$6 minutes).  The (phase + amplitude) self-calibration provided a dramatic improvement in quality.  The typical peak SNR increased by a factor of 3; the resulting noise was 30 $\mu$Jy beam$^{-1}$ (10 mK) for a $\sim$0\farcs25 beam.  The same procedure was applied to archival datasets.  

Next, we prepared the extended data as was described above, with an additional time-averaging to 6 s integrations (from the original 2 s records).  The data for each individual extended observation were imaged and checked for misalignments and flux discrepancies.  Once those are corrected (if necessary; see Section~\ref{sec:align}), the compact (already self-calibrated) and extended datasets were combined.  The phases for this combined dataset were iteratively self-calibrated on solution intervals of \{900, 360, 180, 60, 30 s\} (usually only the latter 3 are necessary).  The SPWs were combined in this case to enhance the SNR on longer baselines.  For antenna pairings with SNR $\le 1.5$ on these intervals, the self-calibration solutions were not applied but the corresponding data were not flagged ({\tt applymode=`calonly'} in the {\tt applycal} task).  The sequence was stopped when the peak SNR does not increase by $> 5$\%\ and the map quality does not visually improve.  One iteration of amplitude self-calibration was attempted on the starting interval of the phase self-calibration sequence.  

\begin{figure*}[t!]
\includegraphics[width=\linewidth]{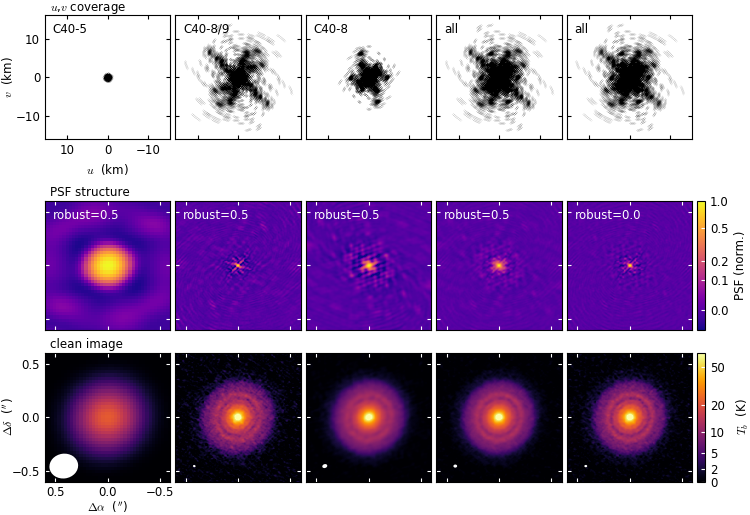}
\figcaption{Illustration of the effects of spatial frequency coverage and visibility weighting on PSF structure and image morphology, for the RU Lup disk.  The top panels show the observed $u$,$v$ coverage.  The middle panels show the corresponding PSF structures, annotated with the corresponding {\tt robust} weighting parameter.  The bottom panels show the corresponding images, on the same $T_b$ scale, created from subsets of the end product of the self-calibration.  FWHM beam dimensions are marked in the lower left corners of each image.    
\label{fig:imaging}
}
\end{figure*}

This self-calibration of the combined datasets resulted in a typical improvement of 40\%\ in the peak SNR, although there is a large range in benefits across the sample.  The improvements are generally smaller here because the compact data were already self-calibrated and the extended data were taken in excellent conditions.  The typical noise measured in the combined, self-calibrated datasets is 10--20 $\mu$Jy beam$^{-1}$ (0.1--0.5 K).            

Once the continuum self-calibration was satisfactory, the same gain tables are applied to the non-spectrally-averaged visibilities (after any required astrometric and flux calibration adjustments) to obtain a corresponding calibrated measurement set for the region of the spectrum around the CO $J$=2$-$1 emission line.

\subsection{Imaging During Self-Calibration} \label{sec:imaging}

Self-calibration uses continuum emission models assembled from the `clean' components derived from interferometric imaging.  We adopted a set of imaging standards to homogenize that process.  These were informed by considerable experimentation with the associated parameter choices.  We explored alternative sets of deconvolution scales, clean thresholds, masks, and pixel sizes and found that reasonable other options had negligible influence on the end products of self-calibration.

All imaging was performed with the {\tt tclean} task.  For the compact data, we imaged out to the primary beam FWHM (26\arcsec) with 30 mas pixels ($\sim$10 per synthesized beam FWHM, $\theta_b$) to check for problematic background sources.  Finding nothing of concern, we used 9\arcsec-wide images with 3 mas pixels (again, $\sim$10 pixels per $\theta_b$) for the combined datasets.  We used the multi-scale, multi-frequency synthesis \dt{(assuming a flat spectrum)} deconvolution mode \citep{cornwell08} with a Briggs {\tt robust}=0.5 weighting scheme.  Elliptical masks were designed to reflect the target geometry (aspect ratio, position angle) and pad the outer reaches of the emission distribution.  The adopted (Gaussian) deconvolution scales are target-dependent, but always include a point-like contribution and scales comparable to $\theta_b$ and 2--$3 \times \theta_b$; additional scales (increasing by factors of 2--3) could be selected up to the mask radius.  The algorithm was halted on thresholds; 3$\times$ the noise early in the self-calibration sequence, and 2$\times$ the noise for the last phase-only step and the amplitude self-calibration.

\begin{deluxetable*}{lccccc|cc|l}
\tabletypesize{\scriptsize}
\tablecaption{DSHARP Fiducial Continuum Image Properties \label{table:images}}
\tablehead{
\colhead{Name} & 
\colhead{$\nu$} & 
\colhead{$\theta_b$, \phn\phn PA$_b$} & 
\colhead{RMS noise} & 
\colhead{peak $I_\nu$, $T_b$} & 
\colhead{$F_\nu$} &
\colhead{{\tt robust}} & 
\colhead{$\theta_{\rm tap}$, \phd PA$_{\rm tap}$} & 
\colhead{Refs.} \\
\colhead{} & 
\colhead{(GHz)} & 
\colhead{(mas, \phn\phn \degr)} & 
\colhead{($\mu$Jy\,beam$^{-1}$, K)} & 
\colhead{(mJy\,beam$^{-1}$, K)} & 
\colhead{(mJy)} & 
\colhead{} & 
\colhead{(mas, \phn\phn \degr)} & 
\colhead{}
}
\colnumbers
\startdata
HT Lup    &  239.0 & $38 \times 33$, \phn61    & 14, \phn\phn 0.24 & 8.25, \phn 140       & \phn77  & \phn\phd0.5 & \nodata                   &  IV       \\
GW Lup    &  239.0 & $45 \times 43$, \phn\phn1 & 15, \phn\phn 0.17 & 3.35, \phn\phn 37    & \phn89  & \phn\phd0.5 & $35 \times 15$, \phn\phn0 &  II       \\
IM Lup    &  239.0 & $44 \times 43$, 115       & 14, \phn\phn 0.16 & 7.11, \phn\phn 80    &  253    & \phn\phd0.5 & $33 \times 26$, 138       &  II, III  \\
RU Lup    &  239.0 & $25 \times 24$, 129       & 21, \phn\phn 0.73 & 3.45, \phn 123       &  203    & $-$0.5      & $22 \times 10$, 174       &  II       \\
Sz 114    &  239.0 & $67 \times 28$, \phn92    & 19, \phn\phn 0.22 & 3.36, \phn\phn 38    & \phn49  & \phn\phd0.5 & \nodata                   &  II       \\
Sz 129    &  239.0 & $44 \times 31$, \phn94    & 15, \phn\phn 0.24 & 0.96, \phn\phn 15    & \phn86  & \phn\phd0.0 & \nodata                   &  II       \\
MY Lup    &  239.0 & $44 \times 43$, 122       & 16, \phn\phn 0.18 & 1.78, \phn\phn 20    & \phn79  & \phn\phd0.0 & $39 \times 15$, 163       &  II       \\
HD 142666 &  231.9 & $32 \times 22$, \phn62    & 13, \phn\phn 0.35 & 1.28, \phn\phn 41    &  130    & \phn\phd0.5 & \nodata                   &  II       \\
HD 143006 &  239.0 & $46 \times 45$, \phn51    & 15, \phn\phn 0.15 & 0.67, \phn\phn\phn 7 & \phn59  & \phn\phd0.0 & $42 \times 20$, 172       &  II, X    \\
AS 205    &  233.7 & $38 \times 25$, \phn95    & 16, \phn\phn 0.38 & 6.15, \phn 145       &  358    & \phn\phd0.5 & \nodata                   &  IV       \\
SR 4      &  239.0 & $34 \times 34$, \phn10    & 25, \phn\phn 0.46 & 3.40, \phn\phn 63    & \phn69  & $-$0.5  & $35 \times 10$, \phn\phn0     &  II       \\
Elias 20  &  239.0 & $32 \times 23$, \phn76    & 15, \phn\phn 0.44 & 2.59, \phn\phn 75    & 104     & \phn\phd0.0 & \nodata                   &  II       \\
DoAr 25   &  239.0 & $41 \times 22$, \phn70    & 13, \phn\phn 0.31 & 1.35, \phn\phn 32    & 246     & \phn\phd0.5 & \nodata                   &  II       \\
Elias 24  &  231.9 & $37 \times 34$, \phn82    & 19, \phn\phn 0.49 & 4.63, \phn 119       & 352     & \phn\phd0.0 & $35 \times 10$, 166       &  II       \\
Elias 27  &  231.9 & $49 \times 47$, \phn47    & 14, \phn\phn 0.14 & 4.83, \phn\phn 48    & 330     & \phn\phd0.5 & $40 \times 20$, 173       &  II, III  \\
DoAr 33   &  239.0 & $37 \times 24$, \phn75    & 17, \phn\phn 0.41 & 1.89, \phn\phn 46    & \phn35  & \phn\phd0.0 & $20 \times 10$, 167       &  II       \\
WSB 52    &  239.0 & $33 \times 27$, \phn74    & 16, \phn\phn 0.38 & 2.60, \phn\phn 62    & \phn67  & \phn\phd0.0 & \nodata                   &  II, III  \\
WaOph 6   &  239.0 & $58 \times 54$, \phn84    & 17, \phn\phn 0.12 & 8.67, \phn\phn 59    & 161     & \phn\phd0.0 & $55 \times 10$, \phn10    &  II, III  \\
AS 209    &  239.0 & $38 \times 36$, \phn68    & 19, \phn\phn 0.30 & 1.83, \phn\phn 29    & 288     & $-$0.5  & $37 \times 10$, 162           &  II, VIII \\
HD 163296 &  239.0 & $48 \times 38$, \phn82    & 23, \phn\phn 0.27 & 4.26, \phn\phn 50    & 715     & $-$0.5  & \nodata                       &  II, IX   \\
\enddata
\tablecomments{Col.~(1) Target name.  Col.~(2) Mean frequency.  Col.~(3) Synthesized beam FWHM and position angle.  Col.~(4) RMS noise in the map, as described in Section~\ref{sec:imaging}.  Col.~(5) Peak intensity in the map.  \dt{Note that noise and peak brightness temperatures are calculated assuming the Rayleigh-Jeans limit.}  Col.~(6) Integrated flux density inside the image mask.  Col.~(7) Briggs {\tt robust} value.  Col.~(8) FWHM and position angle of the taper (if applicable).}
\tablerefs{II = \citet{huang18b}, III = \citet{huang18c}, IV = \citet{troncoso18}, VIII = \citet{guzman18}, IX = \citet{isella18}, X = \citet{perez18}.}  
\end{deluxetable*}

Special effort was made to verify that sidelobes in the point spread function (PSF, or `dirty' beam) do not corrupt the self-calibration.  The extended ALMA configurations place antennas along three distinct arms (set by the site topography).  The corresponding spatial frequency coverage generates complicated PSF features, with sidelobes up to $\sim$30\%.  Figure~\ref{fig:imaging} illustrates the impact, showing the connections between the sampling function ($u$,$v$ coverage), PSF, and image for different configurations and weighting schemes.  We vetted the effects of those PSF features on self-calibration by repeating the process for different combinations of weighting schemes and tapers.  Coupling lower {\tt robust} values with tapers can mitigate PSF artifacts while maintaining resolution, but at a substantial SNR cost.  Direct comparisons (of both visibilities and images) between these variants and the standard methodology outlined above demonstrated that the PSF features had negligible impact on the self-calibration.\footnote{The HD 163296 disk is the one exception (albeit a quite modest one): we find $\sim$10\%\ SNR improvements (relative to the standard) when self-calibration is conducted for images with {\tt robust}=-0.5, due to the combination of the target emission distribution and the unusual spatial frequency coverage from the archival data.}

While the effects on self-calibration are minimal, the resulting images can still exhibit PSF-related artifacts.  One of the more interesting is the imprint of a hexagonal structure on emission rings (e.g., second image from left, bottom row of Figure~\ref{fig:imaging}), produced by convolution with a ``spoked" PSF (a consequence of the the extended ALMA configuration arms).  As demonstrated in the bottom right panel, this can usually be minimized with an appropriate visibility weighting and/or tapering.

\subsection{Astrometric and Flux Scale Alignment} \label{sec:align}

Half the sample targets show clear spatial offsets between \dt{their emission centers in} different observations.  For the larger of these shifts \dt{($\sim$100 mas)}, the cause is proper motion (especially when using archival data); in other cases, smaller (10--30 mas) mismatches might instead be attributed to instrumental or atmospheric artifacts.  Combining these datasets without correcting these shifts creates blurred (or even double) images, which is problematic when they are used as initial self-calibration models.  The solution is to simply adjust the visibility phases to shift into alignment.  We measure emission centroid positions with Gaussian fits in the image plane for each individual observation and calculate the offsets relative to the highest quality extended dataset.  The {\tt fixvis} task then implements the appropriate phase adjustments.  In cases where the observations have different pointing centers, we manually reconcile them with the {\tt fixplanets} task.     

We also routinely found mismatches in the amplitude scales among different observations of a target.  Some experimentation showed that noticeably improved self-calibration results were obtained if the relative flux scales between observations were consistent within 5\%.  To quantify any mismatches, we inspected the deprojected (according to the Gaussian fit geometries noted above), azimuthally-averaged visibilities from different datasets on 200-500 k$\lambda$ baseline lengths (at lower spatial frequencies, the extended configuration data are too sparse, and at higher frequencies the averages are more strongly affected by low SNR and phase noise). 

These mismatches are caused by inaccurate flux calibration.  The claimed calibration accuracy is $\sim$10\%, although the adopted methodology for estimating calibrator fluxes (interpolation in time and frequency) can lead to some added uncertainty.  About a third of the sample had 5--10\%\ mismatches, but the majority exhibited 15--25\%\ discrepancies for at least one dataset.  In some cases, these were tracked down to a bookkeeping issue: the data were pipeline-processed before a relevant calibrator catalog update.  Some 2017 November datasets that used J1427-4206 as the calibrator were problematic.  There is no obvious error in the calibrator catalog, so the issue must be with the interpolation: perhaps this quasar flared or changed its spectrum between catalog entries.  Regardless of the cause, these misalignments were rectified.  We selected a reference dataset and used the {\tt gaincal} task to re-scale the outlier datasets.

\subsection{\dt{Fiducial Images}} \label{sec:fiducial}

\begin{figure*}[htp!]
\centering
\includegraphics[width=6.625in]{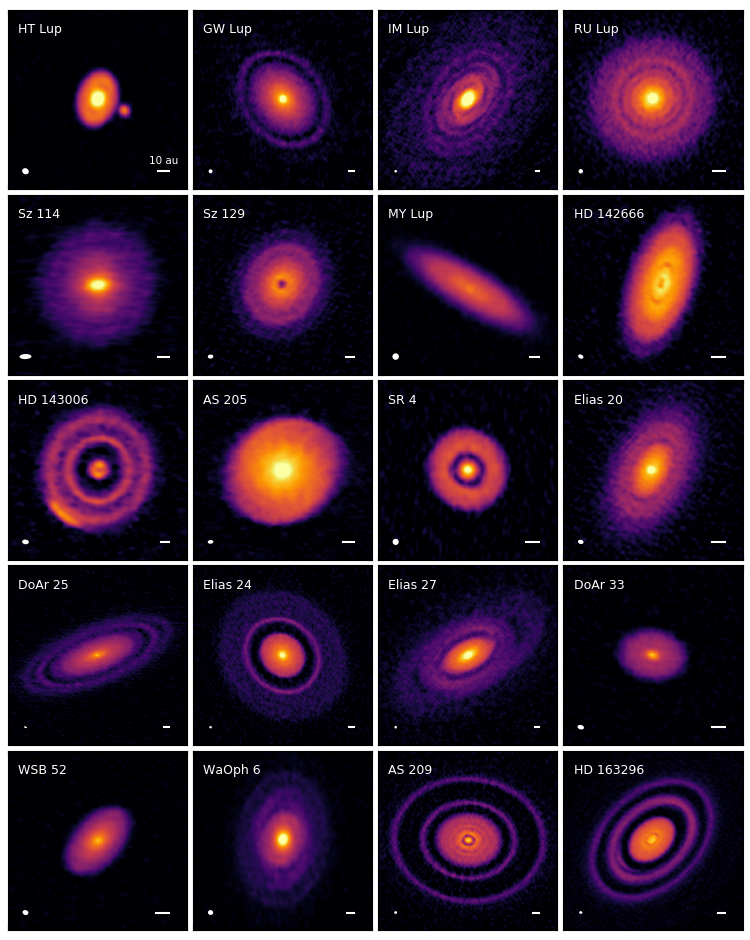}
\figcaption{A gallery of 240 GHz (1.25 mm) continuum emission images for the disks in the DSHARP sample.  Beam sizes and 10 au scalebars are shown in the lower left and right corners of each panel, respectively.  All images are shown with \dt{an asinh stretch to reduce the dynamic range (accentuate fainter details without over-saturating the bright emission peaks)}.  For more quantitative details regarding the image dimensions and intensity scales, see \citet{huang18b} and \citet{troncoso18}.
\label{fig:gallery}
}
\end{figure*}

\begin{figure}
\centering
\includegraphics[width=\linewidth]{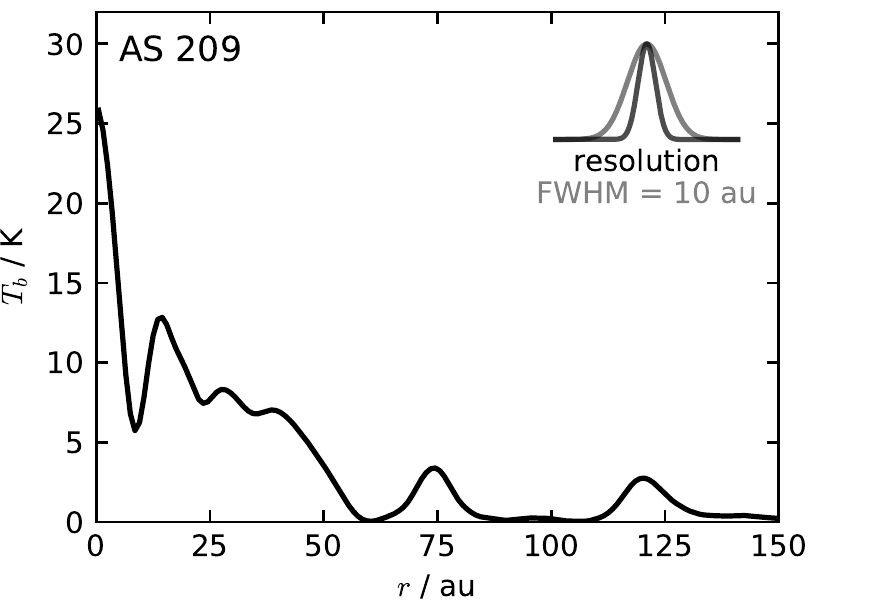}
\figcaption{\dt{The deprojected, azimuthally-averaged radial brightness temperature profile for the 240 GHz continuum emission from the AS 209 disk \citep[see][for more details]{huang18b,guzman18}.  The corresponding image is shown in the bottom row of Figure~\ref{fig:gallery}, second from right.  The PSF profile (resolution) is marked in black in the upper right corner, along with a gray Gaussian profile that has FWHM = 10 au, to illustrate that the disk substructures typically have compact dimensions.}
\label{fig:as209_prof}
}
\end{figure}

After the calibration was complete, we synthesized a set of fiducial images for further analysis.  The continuum imaging followed the methodology outlined in Section~\ref{sec:imaging}, but was tailored to individual sources with the aim of minimizing PSF artifacts.  In many cases, this involved adopting a visibility weighting scheme that traded SNR for resolution, as well as a visibility taper to improve the PSF symmetry.  Table~\ref{table:images} lists the basic parameters and resulting properties of these fiducial images.  A gallery of the continuum images are shown in Figure~\ref{fig:gallery}.  \dt{Small-scale substructures are notable in all of the DSHARP targets, often with compact (FWHM $\lesssim$ 10 au) dimensions.  Figure~\ref{fig:as209_prof} emphasizes the utility of pushing the ALMA resolution for recovering such features in one particularly illustrative example.}    

We also synthesized channel maps of the CO $J$=2$-$1 emission following the basic steps outlined above.  The self-calibrated CO visibilities were continuum-subtracted and imaged in LSRK velocity channels at roughly the native channel spacing (0.35 km s$^{-1}$; the actual velocity resolution is about two channels, due to Hanning smoothing in the ALMA correlator).  The DSHARP data are generally not sensitive enough to reconstruct useful channel maps of the emission line at the best available resolution.  We compromised by increasing the relative weight of shorter baselines and employing a modest taper.  Table~\ref{table:CO} lists the imaging parameters, and Figure~5 shows the channel maps.  \dt{For many of the targets, the CO channel maps exhibit partially recovered large-scale emission structures from the ambient cloud material.  These are noted in Table~\ref{table:CO} to prevent confusion in the interpretation of extended emission features in some cases (e.g., Elias 24 and WSB 52 are particularly problematic cases).}

\section{Data Release} \label{sec:results}

One key inspiration for conducting the DSHARP survey was to provide a set of resources to the community that can seed and develop a range of related work.  To that end, we have released a suite of data products that go beyond the standard contents in the ALMA archive.  This release is available online at \url{https://almascience.org/alma-data/lp/DSHARP}.  It includes: (1) {\tt CASA} scripts and associated {\tt python} modules used to calibrate and image the data; (2) fully calibrated continuum and CO measurement sets (visibility datafiles); (3) continuum images and CO channel maps; and (4) some secondary products (radial intensity profiles, SED data).  With this data release and the standard ALMA archive products, the community has the access needed to both reproduce and expand on the efforts detailed in the initial series of DSHARP articles.

\begin{deluxetable*}{lccc|cc|ll}
\tabletypesize{\scriptsize}
\tablecaption{DSHARP Fiducial CO Datacube Properties \label{table:CO}}
\tablehead{
\colhead{Name} & 
\colhead{$\theta_b$, \phn\phn PA$_b$} & 
\colhead{RMS noise} & 
\colhead{peak $I_\nu$, $T_b$} & 
\colhead{{\tt robust}} & 
\colhead{$\theta_{\rm tap}$, \phd PA$_{\rm tap}$} & 
\colhead{comments} &
\colhead{Refs.} \\
\colhead{} & 
\colhead{(mas, \phn\phn \degr)} & 
\colhead{(mJy\,beam$^{-1}$, K)} & 
\colhead{(mJy\,beam$^{-1}$, K)} & 
\colhead{} & 
\colhead{(mas, \phn\phn \degr)} & 
\colhead{} & 
\colhead{}
}
\colnumbers
\startdata
HT Lup    & $\phn53\times\phn50$, \phn\phn66    & 1.2, \phn 10.1    & 12.8, \phn 111    & 0.5 & \nodata & cloud (severe); $< 150$ m filter  & IV      \\
GW Lup    & $109\times\phn81$, \phn\phn97       & 1.4, \phn\phn 3.5 & 20.9, \phn\phn 54 & 1.0 & $\phn40\times\phn40$, \nodata & \nodata                         & \nodata \\
IM Lup    & $122\times115$, \phn\phn47          & 1.9, \phn\phn 3.2 & 34.4, \phn\phn 56 & 0.0 & $100\times100$, \nodata       & \nodata                         & III     \\
RU Lup    & $\phn95\times\phn83$, \phn\phn72    & 1.2, \phn\phn 3.4 & 49.7, \phn 146    & 1.0 & $\phn40\times\phn40$, \nodata & cloud (mild), complex outflow   & \nodata \\
Sz 114    & $130\times\phn90$, \phn105          & 2.0, \phn\phn 4.0 & 26.8, \phn\phn 53 & 1.0 & $\phn40\times\phn40$, \nodata & cloud (moderate)                & \nodata \\
Sz 129    & $110\times\phn83$, \phn\phn76       & 1.0, \phn\phn 2.6 & 15.3, \phn\phn 38 & 1.0 & $\phn40\times\phn40$, \nodata & \nodata                         & \nodata \\
MY Lup    & $100\times\phn82$, \phn\phn79       & 1.1, \phn\phn 3.1 & 15.4, \phn\phn 43 & 1.0 & $\phn40\times\phn40$, \nodata & cloud (mild)                    & \nodata \\
HD 142666 & $\phn77\times\phn61$, \phn\phn81    & 1.3, \phn\phn 6.3 & 12.4, \phn\phn 61 & 1.0 & $\phn40\times\phn40$, \nodata & \nodata                         & \nodata \\
HD 143006 & $\phn66\times\phn49$, \phn\phn84    & 1.0, \phn\phn 7.1 & 11.2, \phn\phn 80 & 0.8 & $\phn20\times\phn20$, \nodata & \nodata                         & X       \\
AS 205    & $115\times\phn92$, \phn\phn93       & 1.4, \phn\phn 3.2    & 75.9, \phn 166    & 1.0 & $\phn40\times\phn40$, \nodata      &  \nodata                              & IV      \\
SR4       & $111\times\phn87$, \phn\phn90       & 1.5, \phn\phn 3.5 & 29.9, \phn\phn 71 & 1.0 & $\phn40\times\phn40$, \nodata & cloud (moderate)                & \nodata \\
Elias 20  & $102\times\phn72$, \phn\phn88       & 1.8, \phn\phn 5.6 & 27.0, \phn\phn 85 & 1.0 & $\phn40\times\phn40$, \nodata & cloud (severe), outflow         & \nodata \\
DoAr 25   & $101\times\phn78$, \phn\phn87       & 1.3, \phn\phn 3.9 & 18.9, \phn\phn 55 & 1.0 & $\phn40\times\phn40$, \nodata & cloud (moderate)                & \nodata \\
Elias 24  & $\phn94\times\phn60$, \phn\phn90    & 1.5, \phn\phn 6.2 & 23.6, \phn\phn 97 & 1.0 & $\phn40\times\phn40$, \nodata & cloud (severe)                  & \nodata \\
Elias 27  & $132\times111$, \phn123             & 1.6, \phn\phn 2.5 & 44.9, \phn\phn 71 & 1.0 & $100\times\phn70$, \phn145    & cloud (moderate), envelope$?$   & III     \\
DoAr 33   & $103\times\phn79$, \phn\phn88       & 1.3, \phn\phn 3.6 & 15.8, \phn\phn 45 & 1.0 & $\phn40\times\phn40$, \nodata & cloud (mild)                    & \nodata \\
WSB 52    & $114\times\phn80$, \phn\phn90       & 1.1, \phn\phn 2.8 & 31.4, \phn\phn 79 & 1.0 & $\phn40\times\phn40$, \nodata & cloud (severe), complex outflow & \nodata \\
WaOph 6   & $126\times115$, \phn100             & 1.3, \phn\phn 2.1 & 41.0, \phn\phn 65 & 0.5 & $100\times\phn30$, \phn\phn17 & cloud (mild)                    & III     \\
AS 209    & $\phn95\times\phn72$, \phn\phn96    & 0.9, \phn\phn 2.9 & 21.4, \phn\phn 72 & 1.0 & $\phn25\times\phn10$, \phn\phn10       & cloud (mild)                    & VIII    \\
HD 163296 & $104\times\phn95$, \phn100          & 0.8, \phn\phn 1.9 & 40.7, \phn\phn 95 & 0.5 & \nodata                       & \nodata                         & IX      \\
\enddata
\tablecomments{Col.~(1) Target name.  Col.~(2) Synthesized beam FWHM and position angle.  Col.~(3) RMS noise per channel, measured as described in Section~\ref{sec:imaging}.  HD 143006 and HD 163296 are imaged with 0.32 km s$^{-1}$ channels; for all other targets, we used 0.35 km s$^{-1}$ channels.  Col.~(4) Peak intensity.  \dt{Note that noise and peak brightness temperatures are calculated assuming the Rayleigh-Jeans limit.}  Col.~(5) Briggs {\tt robust} value.  Col.~(6) FWHM and position angle of the adopted taper (if applicable).  Col.~(7) Comments on issues with the channel maps, including degree of contamination from the ambient molecular cloud and the presence of non-disk features.}
\tablerefs{III \citet{huang18c}. IV \citet{troncoso18}. VIII \citet{guzman18}.  IX \citet{isella18}.  X \citet{perez18}.}  
\end{deluxetable*}

\section{Overview: Initial DSHARP Results} \label{sec:discussion}

This article has detailed the scientific motivations behind DSHARP, introduced the survey strategy and sample, described the observations and calibration process, and presented the resulting products as part of our data release.  It is also the first in a series of articles that explore and analyze the data in more detail.  The principal DSHARP conclusions can be summarized as follows: 

\medskip

$\bullet$ Continuum substructures are ubiquitous \dt{in this sample}, as can be deduced from Figure~\ref{fig:gallery}.  Small-scale emission features are found at effectively any disk radius, from 5 au out to more than 150 au.
    
\medskip
    
$\bullet$ The most common form of these substructures are concentric bright rings and dark gaps.  There are no obvious patterns in their distributions or connections to the stellar host properties.  There are hints of ring/gap substructures that are obfuscated due to their smaller size scales (relative to the DSHARP resolution) and/or their modest amplitudes with respect to an optically thick background in the inner disk.  Measurements of the rings and gaps, as well as a more detailed exploration of \dt{their potential origins} and associated issues, are presented by \citet{huang18b}.   
    
\medskip
    
$\bullet$ While less common, the spiral morphologies identified for a subset of disks in the DSHARP sample are striking.  For the cases with apparently single host stars (IM Lup, Elias 27, and WaOph 6), the spiral patterns are complex and appear to be superposed with rings and gaps.  Their emission distributions and potential origins are characterized by \citet{huang18c}.

\medskip
    
$\bullet$ For the two known multiple star systems in the DSHARP sample, HT Lup and AS 205, the disks around the primary stars show clear two-armed spirals and complicated CO distributions that are indicative of strong dynamical interactions.  The circumstellar material in these systems is studied by \citet{troncoso18}.

\medskip
    
$\bullet$ Azimuthal asymmetries are rare in this sample.  Substantial deviations from axisymmetry (or point symmetry for the spirals) are only identified in two cases.  The disks around HD 143006 and HD 163296 show small, arc-shaped features in otherwise emission-depleted regions (i.e., beyond the continuum disk edge and in a gap, respectively).  The properties and potential origins of these special cases are scrutinized by \citet{perez18} and \citet{isella18}, respectively.    

\medskip
    
$\bullet$ In some cases, the continuum emission can be decomposed into {\it only} small-scale substructures.  The AS 209 disk is a particularly compelling example.  \citet{guzman18} quantify its substructures and highlight an important point: there are analogous features lurking in the gas (even as traced by optically thick $^{12}$CO), at radii well beyond the extent of the continuum emission.

\medskip
    
$\bullet$ The ring substructure sizes and amplitudes suggest that these features can be understood as dust trapped in axisymmetric gas pressure bumps.  \citet{dullemond18b} demonstrate this conclusion and derive a lower limit on the strength of the turbulence in the disk.  These and other related analyses are guided by a fiducial dust model developed by \citet{birnstiel18}.    

\medskip

$\bullet$ A new suite of hydrodynamics simulations by \citet{zhang18} suggest that dynamical interactions between low-mass (sub-Jupiter) planets and their local disk material are plausible explanations of the observed ring/gap substructures.  Assuming this is the case, those simulations are used to reconstruct the associated planet population in the mass -- semimajor axis plane.
    
\bigskip

There is, of course, much more to learn from the DSHARP dataset.  Our hope is that this preliminary foray not only provides useful results and motivation for many other studies, but also lays some technical groundwork for designing and calibrating future ALMA surveys of disks at very high angular resolution.

\acknowledgments We thank Erica Keller and Tony Remijan for their pipeline calibration efforts, the ALMA staff for pushing through the observations even after the re-configuration delays, Todd Hunter and Crystal Brogan for some useful technical consultations, Ian Czekala for his advice on measuring stellar parameters, and an anonymous reviewer for thoughtful suggestions.  S.A. and J.H. acknowledge support from the National Aeronautics and Space Administration under grant No.~17-XRP17$\_$2-0012 issued through the Exoplanets Research Program.  J.H. acknowledges support from the National Science Foundation Graduate Research Fellowship under Grant No.~DGE-1144152.  L.P. acknowledges support from CONICYT project Basal AFB-170002 and from FCFM/U.~de Chile Fondo de Instalaci\'on Acad\'emica.  A.I. acknowledges support from the National Aeronautics and Space Administration under grant No.~NNX15AB06G issued through the Origins of Solar Systems program, and from the National Science Foundation under grant No.~AST-1715719.  C.P.D. acknowledges support by the German Science Foundation (DFG) Research Unit FOR 2634, grants DU 414/22-1 and DU 414/23-1.  V.V.G. and J.C acknowledge support from the National Aeronautics and Space Administration under grant No.~15XRP15$\_$20140 issued through the Exoplanets Research Program.  T.B. acknowledges funding from the European Research Council (ERC) under the European Union’s Horizon 2020 research and innovation programme under grant agreement No.~714769. M.B. acknowledges funding from ANR of France under contract number ANR-16-CE31-0013 (Planet Forming disks).  Z.Z. and S.Z. acknowledge support from the National Aeronautics and Space Administration through the Astrophysics Theory Program with Grant No.~NNX17AK40G and the Sloan Research Fellowship.  L.R. acknowledges support from the ngVLA Community Studies program, coordinated by the National Radio Astronomy Observatory, which is a facility of the National Science Foundation operated under cooperative agreement by Associated Universities, Inc.  This paper makes use of the following ALMA data: ADS/JAO.ALMA \#2016.1.00484.L, ADS/JAO.ALMA \#2011.0.00531.S, ADS/JAO.ALMA \#2012.1.00694.S, ADS/JAO.ALMA \#2013.1.00226.S, ADS/JAO.ALMA \#2013.1.00366.S, ADS/JAO.ALMA \#2013.1.00498.S, ADS/JAO.ALMA \#2013.1.00631.S, ADS/JAO.ALMA \#2013.1.00798.S, ADS/JAO.ALMA \#2015.1.00486.S, ADS/JAO.ALMA \#2015.1.00964.S. ALMA is a partnership of ESO (representing its member states), NSF (USA) and NINS (Japan), together with NRC (Canada), MOST and ASIAA (Taiwan), and KASI (Republic of Korea), in cooperation with the Republic of Chile. The Joint ALMA Observatory is operated by ESO, AUI/NRAO and NAOJ.  This work presents results from the European Space Agency (ESA) space mission {\it Gaia}. {\it Gaia} data are being processed by the {\it Gaia} Data Processing and Analysis Consortium (DPAC). Funding for the DPAC is provided by national institutions, in particular the institutions participating in the {\it Gaia} MultiLateral Agreement (MLA). The {\it Gaia} mission website is \url{https://www.cosmos.esa.int/gaia}. The {\it Gaia} archive website is \url{https://archives.esac.esa.int/gaia}.

\facilities{ALMA}

\software{
{\tt CASA} \citep{mcmullin07}, 
{\tt Numpy} \citep{numpy}, 
{\tt Matplotlib} \citep{matplotlib}, 
{\tt Astropy} \citep{astropy}, 
{\tt ScottiePippen} \citep[][\url{https://github.com/iancze/ScottiePippen}]{czekala16}.
}

\begin{figure*}
\figsetstart
\figsetnum{5}
\figsettitle{DSHARP CO $J$=2$-$1 Channel Maps}

\figsetgrpstart
\figsetgrpnum{5.1}
\figsetgrptitle{HT Lup CO channel maps}
\figsetplot{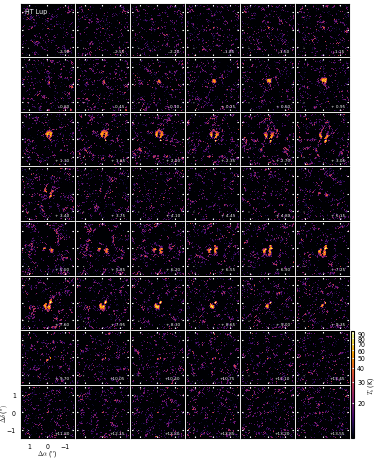}
\figsetgrpnote{Channel maps of the $^{12}$CO $J$=2$-$1 line emission from the HT Lup disk.  Beam dimensions are shown in the lower left corner of each panel.  The LSRK velocity is marked in the lower right corner of each panel.}
\figsetgrpend
 
\figsetgrpstart
\figsetgrpnum{5.2}
\figsetgrptitle{GW Lup CO channel maps}
\figsetplot{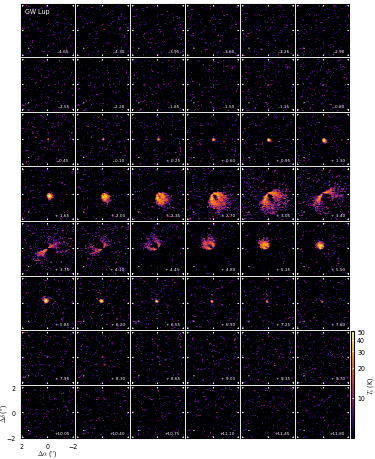}
\figsetgrpnote{Channel maps of the $^{12}$CO $J$=2$-$1 line emission from the GW Lup disk.  Beam dimensions are shown in the lower left corner of each panel.  The LSRK velocity is marked in the lower right corner of each panel.}
\figsetgrpend

\figsetgrpstart
\figsetgrpnum{5.3}
\figsetgrptitle{IM Lup CO channel maps}
\figsetplot{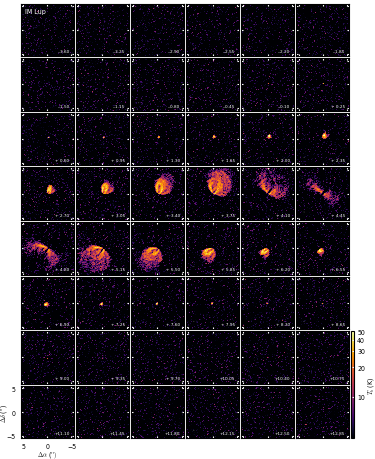}
\figsetgrpnote{Channel maps of the $^{12}$CO $J$=2$-$1 line emission from the IM Lup disk.  Beam dimensions are shown in the lower left corner of each panel.  The LSRK velocity is marked in the lower right corner of each panel.}
\figsetgrpend

\figsetgrpstart
\figsetgrpnum{5.4}
\figsetgrptitle{RU Lup CO channel maps}
\figsetplot{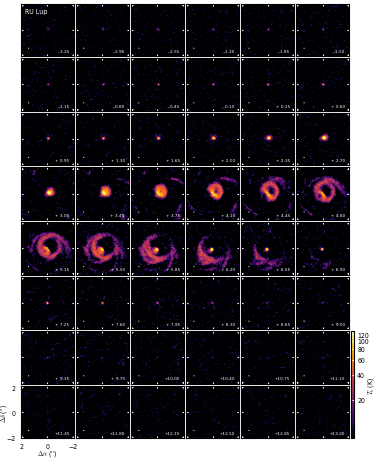}
\figsetgrpnote{Channel maps of the $^{12}$CO $J$=2$-$1 line emission from the RU Lup disk.  Beam dimensions are shown in the lower left corner of each panel.  The LSRK velocity is marked in the lower right corner of each panel.}
\figsetgrpend

\figsetgrpstart
\figsetgrpnum{5.5}
\figsetgrptitle{Sz 114 CO channel maps}
\figsetplot{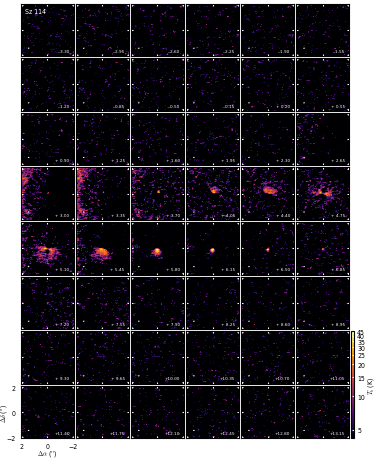}
\figsetgrpnote{Channel maps of the $^{12}$CO $J$=2$-$1 line emission from the Sz 114 disk.  Beam dimensions are shown in the lower left corner of each panel.  The LSRK velocity is marked in the lower right corner of each panel.}
\figsetgrpend

\figsetgrpstart
\figsetgrpnum{5.6}
\figsetgrptitle{Sz 129 CO channel maps}
\figsetplot{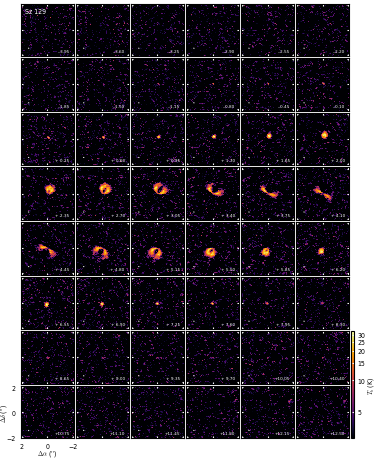}
\figsetgrpnote{Channel maps of the $^{12}$CO $J$=2$-$1 line emission from the Sz 129 disk.  Beam dimensions are shown in the lower left corner of each panel.  The LSRK velocity is marked in the lower right corner of each panel.}
\figsetgrpend

\figsetgrpstart
\figsetgrpnum{5.7}
\figsetgrptitle{MY Lup CO channel maps}
\figsetplot{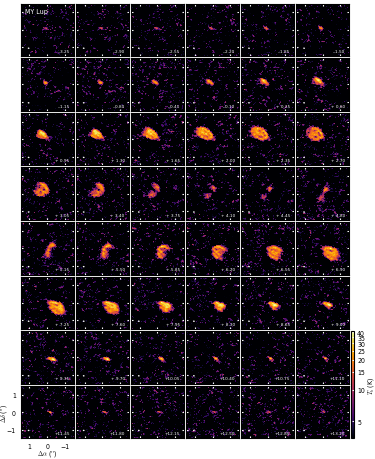}
\figsetgrpnote{Channel maps of the $^{12}$CO $J$=2$-$1 line emission from the Sz 129 disk.  Beam dimensions are shown in the lower left corner of each panel.  The LSRK velocity is marked in the lower right corner of each panel.}
\figsetgrpend

\figsetgrpstart
\figsetgrpnum{5.8}
\figsetgrptitle{HD 142666 CO channel maps}
\figsetplot{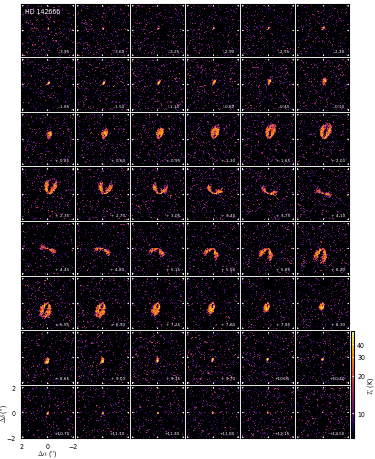}
\figsetgrpnote{Channel maps of the $^{12}$CO $J$=2$-$1 line emission from the HD 142666 disk.  Beam dimensions are shown in the lower left corner of each panel.  The LSRK velocity is marked in the lower right corner of each panel.}
\figsetgrpend

\figsetgrpstart
\figsetgrpnum{5.9}
\figsetgrptitle{HD 143006 CO channel maps}
\figsetplot{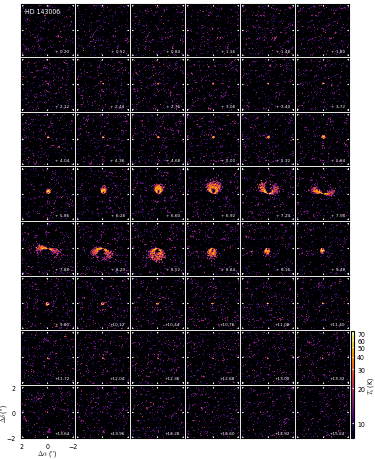}
\figsetgrpnote{Channel maps of the $^{12}$CO $J$=2$-$1 line emission from the HD 143006 disk.  Beam dimensions are shown in the lower left corner of each panel.  The LSRK velocity is marked in the lower right corner of each panel.}
\figsetgrpend

\figsetgrpstart
\figsetgrpnum{5.10}
\figsetgrptitle{AS 205 CO channel maps}
\figsetplot{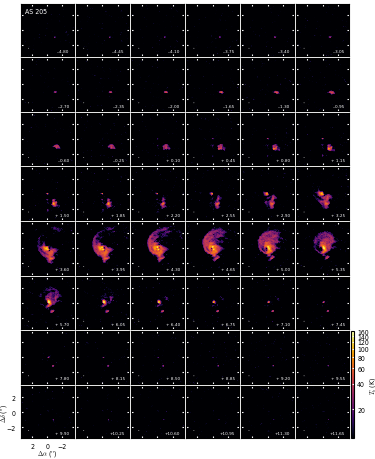}
\figsetgrpnote{Channel maps of the $^{12}$CO $J$=2$-$1 line emission from the AS 205 disk.  Beam dimensions are shown in the lower left corner of each panel.  The LSRK velocity is marked in the lower right corner of each panel.}
\figsetgrpend

\figsetgrpstart
\figsetgrpnum{5.11}
\figsetgrptitle{SR 4 CO channel maps}
\figsetplot{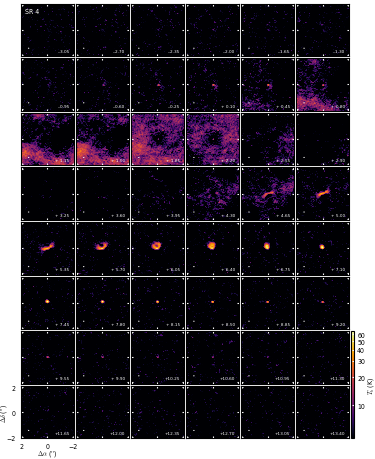}
\figsetgrpnote{Channel maps of the $^{12}$CO $J$=2$-$1 line emission from the SR 4 disk.  Beam dimensions are shown in the lower left corner of each panel.  The LSRK velocity is marked in the lower right corner of each panel.}
\figsetgrpend

\figsetgrpstart
\figsetgrpnum{5.12}
\figsetgrptitle{Elias 20 CO channel maps}
\figsetplot{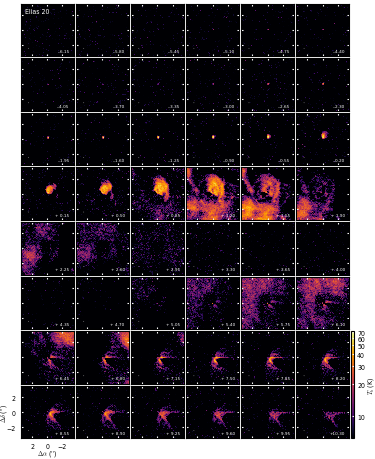}
\figsetgrpnote{Channel maps of the $^{12}$CO $J$=2$-$1 line emission from the Elias 20 disk.  Beam dimensions are shown in the lower left corner of each panel.  The LSRK velocity is marked in the lower right corner of each panel.}
\figsetgrpend

\figsetgrpstart
\figsetgrpnum{5.13}
\figsetgrptitle{DoAr 25 CO channel maps}
\figsetplot{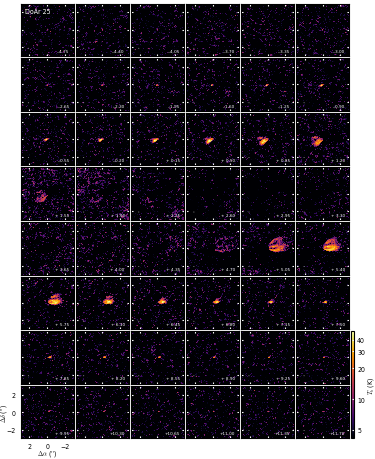}
\figsetgrpnote{Channel maps of the $^{12}$CO $J$=2$-$1 line emission from the DoAr 25 disk.  Beam dimensions are shown in the lower left corner of each panel.  The LSRK velocity is marked in the lower right corner of each panel.}
\figsetgrpend

\figsetgrpstart
\figsetgrpnum{5.14}
\figsetgrptitle{Elias 24 CO channel maps}
\figsetplot{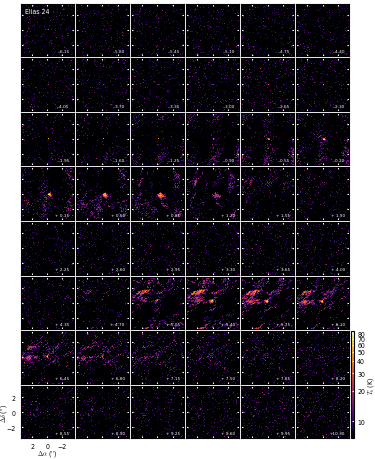}
\figsetgrpnote{Channel maps of the $^{12}$CO $J$=2$-$1 line emission from the Elias 24 disk.  Beam dimensions are shown in the lower left corner of each panel.  The LSRK velocity is marked in the lower right corner of each panel.}
\figsetgrpend

\figsetgrpstart
\figsetgrpnum{5.15}
\figsetgrptitle{Elias 27 CO channel maps}
\figsetplot{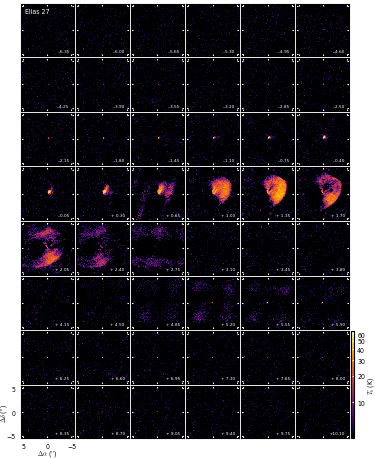}
\figsetgrpnote{Channel maps of the $^{12}$CO $J$=2$-$1 line emission from the Elias 27 disk.  Beam dimensions are shown in the lower left corner of each panel.  The LSRK velocity is marked in the lower right corner of each panel.}
\figsetgrpend

\figsetgrpstart
\figsetgrpnum{5.16}
\figsetgrptitle{DoAr 33 CO channel maps}
\figsetplot{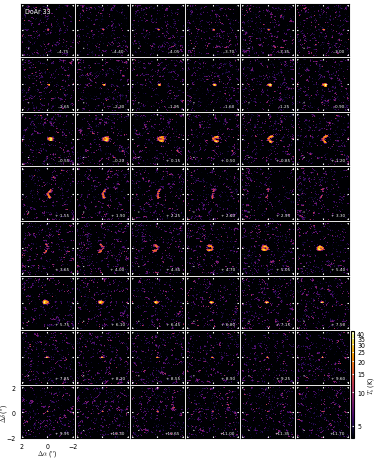}
\figsetgrpnote{Channel maps of the $^{12}$CO $J$=2$-$1 line emission from the DoAr 33 disk.  Beam dimensions are shown in the lower left corner of each panel.  The LSRK velocity is marked in the lower right corner of each panel.}
\figsetgrpend

\figsetgrpstart
\figsetgrpnum{5.17}
\figsetgrptitle{WSB 52 CO channel maps}
\figsetplot{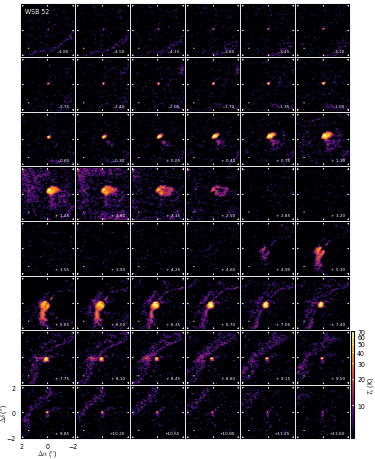}
\figsetgrpnote{Channel maps of the $^{12}$CO $J$=2$-$1 line emission from the WSB 52 disk.  Beam dimensions are shown in the lower left corner of each panel.  The LSRK velocity is marked in the lower right corner of each panel.}
\figsetgrpend

\figsetgrpstart
\figsetgrpnum{5.18}
\figsetgrptitle{WaOph 6 CO channel maps}
\figsetplot{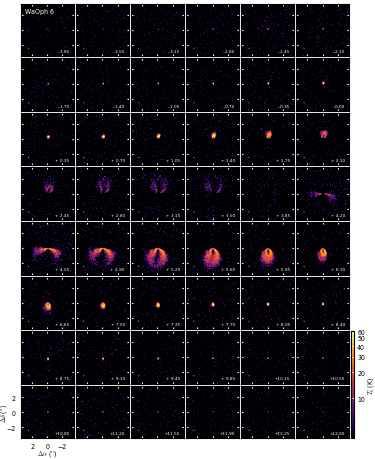}
\figsetgrpnote{Channel maps of the $^{12}$CO $J$=2$-$1 line emission from the WaOph 6 disk.  Beam dimensions are shown in the lower left corner of each panel.  The LSRK velocity is marked in the lower right corner of each panel.}
\figsetgrpend

\figsetgrpstart
\figsetgrpnum{5.19}
\figsetgrptitle{AS 209 CO channel maps}
\figsetplot{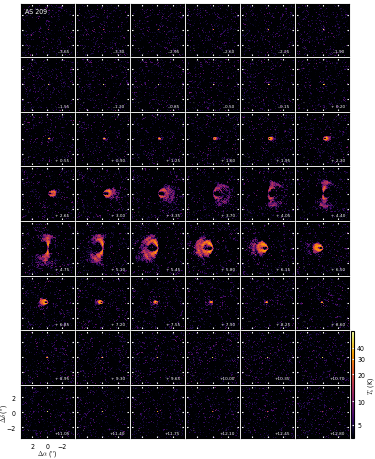}
\figsetgrpnote{Channel maps of the $^{12}$CO $J$=2$-$1 line emission from the AS 209 disk.  Beam dimensions are shown in the lower left corner of each panel.  The LSRK velocity is marked in the lower right corner of each panel.}
\figsetgrpend

\figsetgrpstart
\figsetgrpnum{5.20}
\figsetgrptitle{HD 163296 CO channel maps}
\figsetplot{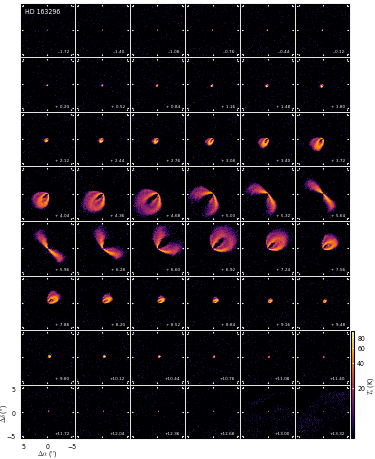}
\figsetgrpnote{Channel maps of the $^{12}$CO $J$=2$-$1 line emission from the HD 163296 disk.  Beam dimensions are shown in the lower left corner of each panel.  The LSRK velocity is marked in the lower right corner of each panel.}
\figsetgrpend
 
\figsetend

\label{figset:chanmaps}

\end{figure*}

\begin{figure}
\figurenum{5.1}
\centering
\includegraphics[width=0.9\linewidth]{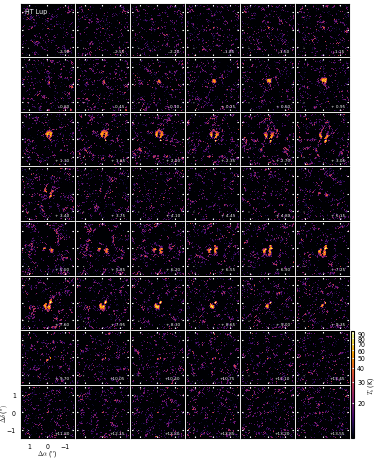}
\figcaption{Channel maps of the $^{12}$CO $J$=2$-$1 line emission from the HT Lup disk.  
}
\end{figure}

\begin{figure}
\figurenum{5.2}
\centering
\includegraphics[width=0.9\linewidth]{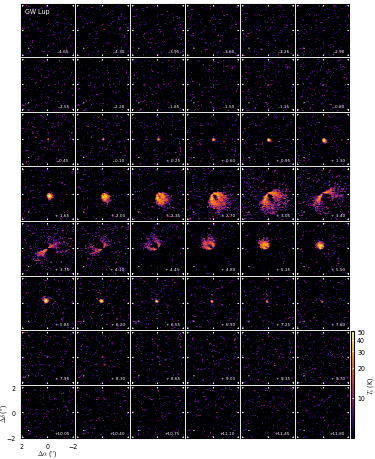}
\figcaption{Channel maps of the $^{12}$CO $J$=2$-$1 line emission from the GW Lup disk.  
}
\end{figure}

\begin{figure}
\figurenum{5.3}
\centering
\includegraphics[width=0.9\linewidth]{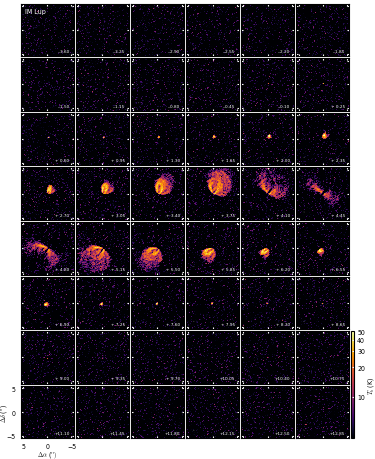}
\figcaption{Channel maps of the $^{12}$CO $J$=2$-$1 line emission from the IM Lup disk.  
}
\end{figure}

\begin{figure}
\figurenum{5.4}
\centering
\includegraphics[width=0.9\linewidth]{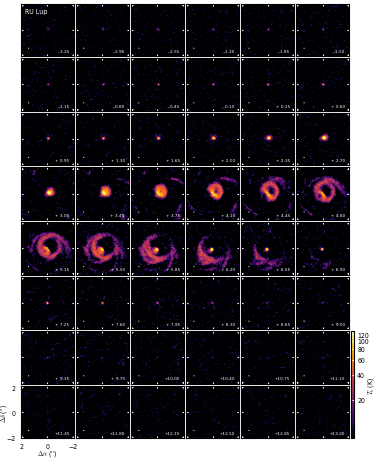}
\figcaption{Channel maps of the $^{12}$CO $J$=2$-$1 line emission from the RU Lup disk.  
}
\end{figure}

\begin{figure}
\figurenum{5.5}
\centering
\includegraphics[width=0.9\linewidth]{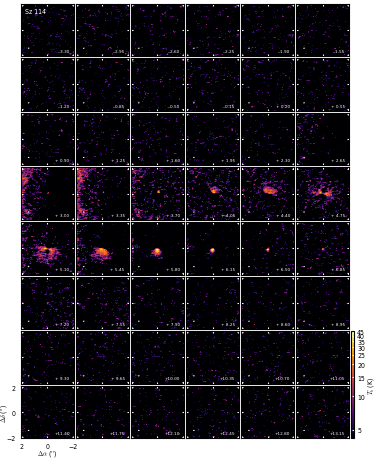}
\figcaption{Channel maps of the $^{12}$CO $J$=2$-$1 line emission from the Sz 114 disk.  
}
\end{figure}

\begin{figure}
\figurenum{5.6}
\centering
\includegraphics[width=0.9\linewidth]{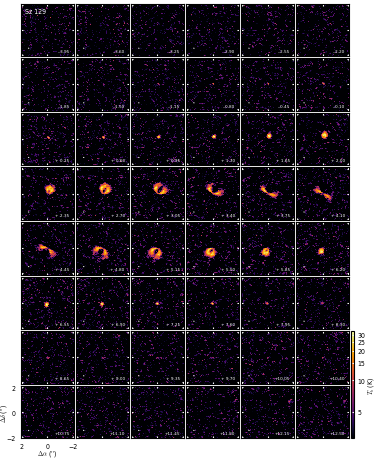}
\figcaption{Channel maps of the $^{12}$CO $J$=2$-$1 line emission from the Sz 129 disk.  
}
\end{figure}

\begin{figure}
\figurenum{5.7}
\centering
\includegraphics[width=0.9\linewidth]{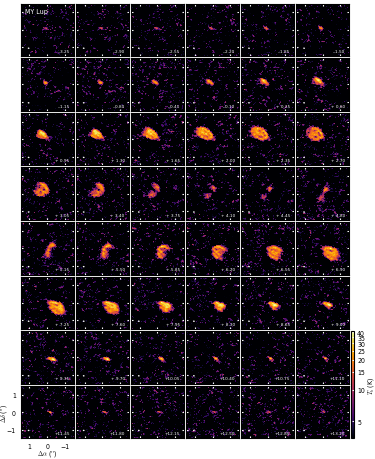}
\figcaption{Channel maps of the $^{12}$CO $J$=2$-$1 line emission from the MY Lup disk. 
}
\end{figure}

\begin{figure}
\figurenum{5.8}
\centering
\includegraphics[width=0.9\linewidth]{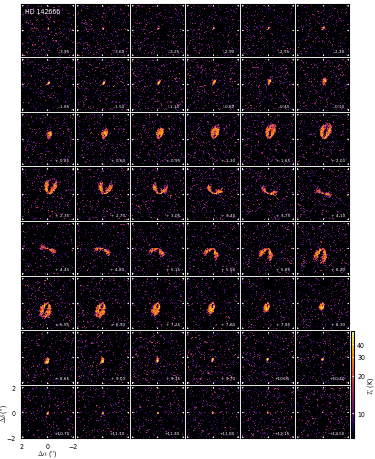}
\figcaption{Channel maps of the $^{12}$CO $J$=2$-$1 line emission from the HD 142666 disk.  
}
\end{figure}

\begin{figure}
\figurenum{5.9}
\centering
\includegraphics[width=0.9\linewidth]{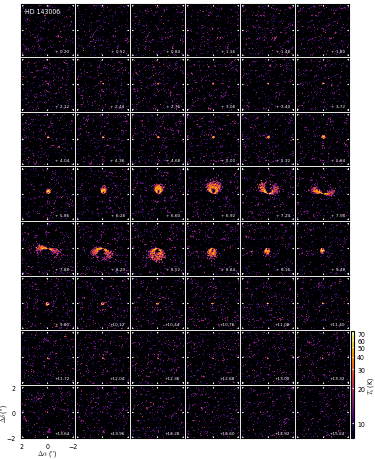}
\figcaption{Channel maps of the $^{12}$CO $J$=2$-$1 line emission from the HD 143006 disk.  
}
\end{figure}

\begin{figure}
\figurenum{5.10}
\centering
\includegraphics[width=0.9\linewidth]{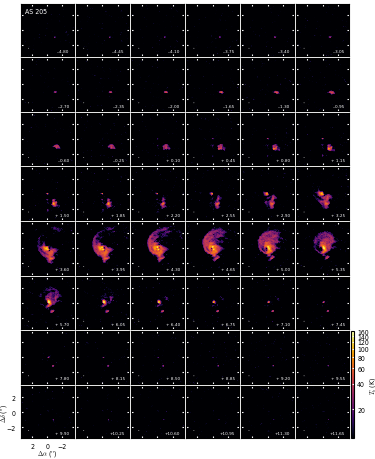}
\figcaption{Channel maps of the $^{12}$CO $J$=2$-$1 line emission from the AS 205 disk.  
}
\end{figure}

\begin{figure}
\figurenum{5.11}
\centering
\includegraphics[width=0.9\linewidth]{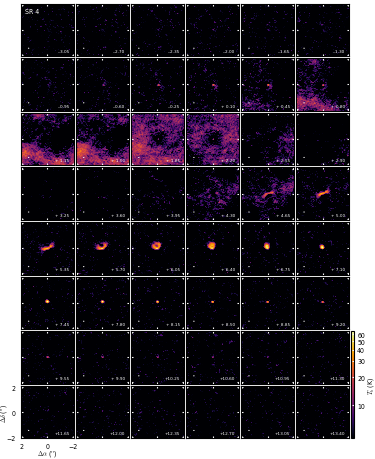}
\figcaption{Channel maps of the $^{12}$CO $J$=2$-$1 line emission from the SR 4 disk.  
}
\end{figure}

\begin{figure}
\figurenum{5.12}
\centering
\includegraphics[width=0.9\linewidth]{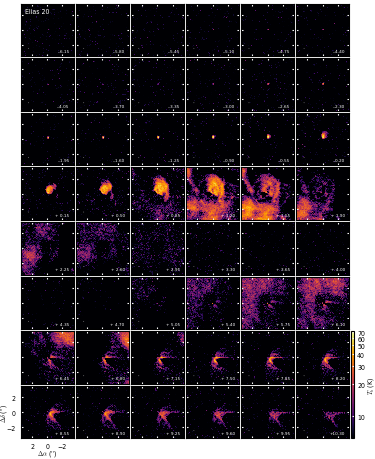}
\figcaption{Channel maps of the $^{12}$CO $J$=2$-$1 line emission from the Elias 20 disk.  
}
\end{figure}

\begin{figure}
\figurenum{5.13}
\centering
\includegraphics[width=0.9\linewidth]{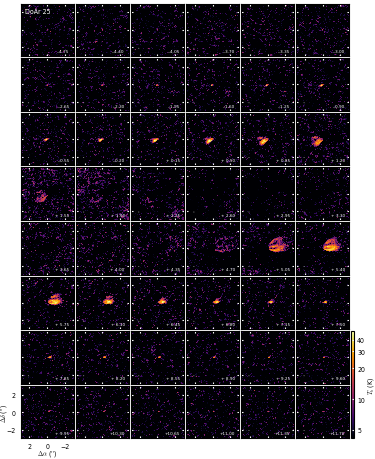}
\figcaption{Channel maps of the $^{12}$CO $J$=2$-$1 line emission from the DoAr 25 disk.  
}
\end{figure}

\begin{figure}
\figurenum{5.14}
\centering
\includegraphics[width=0.9\linewidth]{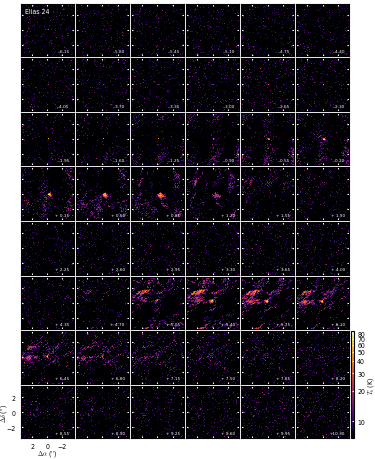}
\figcaption{Channel maps of the $^{12}$CO $J$=2$-$1 line emission from the Elias 24 disk.  
}
\end{figure}

\begin{figure}
\figurenum{5.15}
\centering
\includegraphics[width=0.9\linewidth]{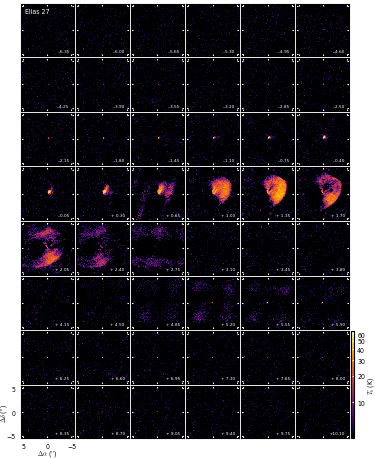}
\figcaption{Channel maps of the $^{12}$CO $J$=2$-$1 line emission from the Elias 27 disk.  
}
\end{figure}

\begin{figure}
\figurenum{5.16}
\centering
\includegraphics[width=0.9\linewidth]{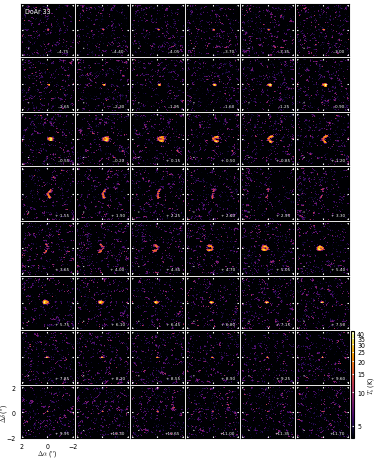}
\figcaption{Channel maps of the $^{12}$CO $J$=2$-$1 line emission from the DoAr 33 disk.  
}
\end{figure}

\begin{figure}
\figurenum{5.17}
\centering
\includegraphics[width=0.9\linewidth]{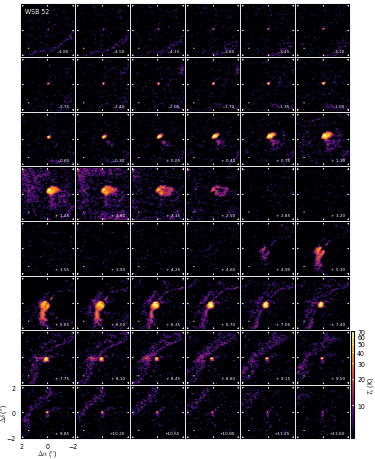}
\figcaption{Channel maps of the $^{12}$CO $J$=2$-$1 line emission from the WSB 52 disk.  
}
\end{figure}

\begin{figure}
\figurenum{5.18}
\centering
\includegraphics[width=0.9\linewidth]{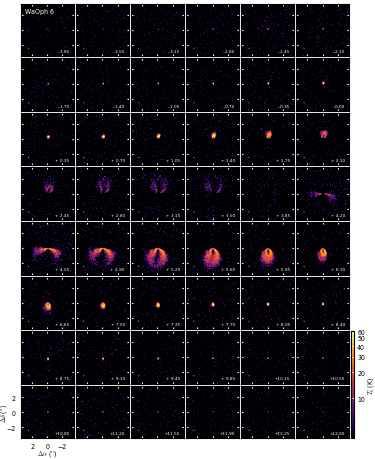}
\figcaption{Channel maps of the $^{12}$CO $J$=2$-$1 line emission from the WaOph 6 disk.  
}
\end{figure}

\begin{figure}
\figurenum{5.19}
\centering
\includegraphics[width=0.9\linewidth]{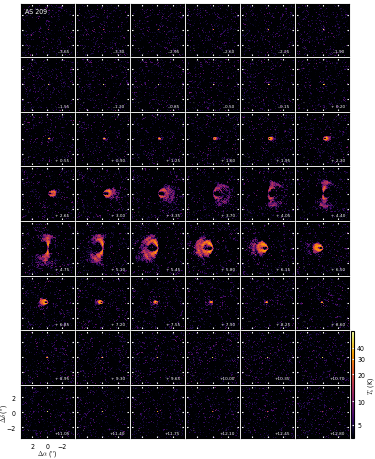} 
\figcaption{Channel maps of the $^{12}$CO $J$=2$-$1 line emission from the AS 209 disk.  
}
\end{figure}

\begin{figure}
\figurenum{5.20}
\centering
\includegraphics[width=0.9\linewidth]{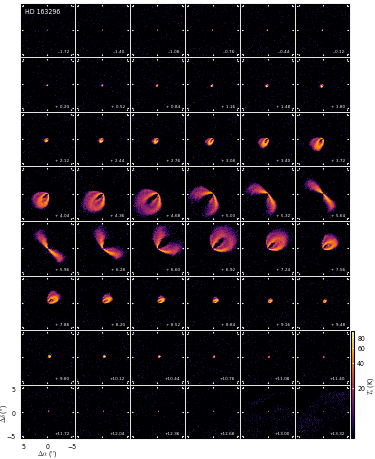}
\figcaption{Channel maps of the $^{12}$CO $J$=2$-$1 line emission from the HD 163296 disk.  
}
\end{figure}

\clearpage
 
\bibliography{references}

\clearpage

\startlongtable
\begin{deluxetable*}{lclccccc}
\tabletypesize{\scriptsize}
\tablenum{2}
\tablecaption{DSHARP Observing Log (ALMA Program 2016.1.00484.L) \label{table:obs_full}}
\tablehead{
\colhead{Name} & 
\colhead{UTC Date} & 
\colhead{Config.} & 
\colhead{Baselines} & 
\colhead{$N_{\rm ant}$} & 
\colhead{$\mathcal{E} / \degr$} & 
\colhead{PWV/mm} & 
\colhead{Calibrators}
}
\colnumbers
\startdata
HT Lup    & 2017/05/14--04:11 & C40-5   & \phn15\,m -- \phn1.1\,km & 43 & 76--77 & 1.00--1.15 
          & J1517-2422, J1427-4206, J1610-3958, J1540-3906 \\
          & 2017/05/17--02:12 & C40-5   & \phn15\,m -- \phn1.1\,km & 49 & 58--67 & 0.90--1.05
          & J1517-2422, J1517-2422, J1610-3958, J1540-3906 \\
          & 2017/09/24--17:39 & C40-8/9 & \phn41\,m -- 12.1\,km    & 39 & 59--70 & 0.60--1.15
          & J1517-2422, J1517-2422, J1534-3526, J1536-3151 \\
          & 2017/09/24--19:12 & C40-8/9 & \phn41\,m -- 12.1\,km    & 39 & 75--78 & 0.65--1.05 
          & J1517-2422, J1427-4206, J1534-3526, J1536-3151 \\
GW Lup    & 2017/05/14--04:11 & C40-5   & \phn15\,m -- \phn1.1\,km & 43 & 69--72 & 1.00--1.15 
          & J1517-2422, J1427-4206, J1610-3958, J1540-3906 \\
          & 2017/05/17--02:12 & C40-5   & \phn15\,m -- \phn1.1\,km & 49 & 51--61 & 0.90--1.05
          & J1517-2422, J1517-2422, J1610-3958, J1540-3906 \\
          & 2017/09/24--23:31 & C40-8/9 & \phn41\,m -- 14.9\,km    & 40 & 29--39 & 0.80--1.20
          & J1617-5848, J1733-1304, J1534-3526, J1536-3151 \\
          & 2017/11/04--14:59 & C43-9   & 113\,m -- 13.9\,km       & 43 & 58--69 & 0.60--0.80
          & J1427-4206, J1427-4206, J1534-3526, J1536-3151 \\
IM Lup    & 2017/09/25--20:18 & C40-8/9 & \phn41\,m -- 14.9\,km    & 42 & 67--74 & 0.85--1.05
          & J1517-2422, J1517-2422, J1610-3958, J1604-4228 \\
          & 2017/10/24--18:09 & C43-9   & \phn41\,m -- 13.9\,km    & 46 & 69--75 & 0.68--0.78
          & J1517-2422, J1427-4206, J1610-3958, J1604-4228 \\
RU Lup    & 2017/05/14--04:11 & C40-5   & \phn15\,m -- \phn1.1\,km & 43 & 76--78 & 1.00--1.15 
          & J1517-2422, J1427-4206, J1610-3958, J1540-3906 \\
          & 2017/05/17--02:12 & C40-5   & \phn15\,m -- \phn1.1\,km & 49 & 58--68 & 0.90--1.05
          & J1517-2422, J1517-2422, J1610-3958, J1540-3906 \\
          & 2017/09/29--20:46 & C40-8/9 & \phn41\,m -- 15.0\,km    & 43 & 60--70 & 1.25--1.65
          & J1517-2422, J1517-2422, J1610-3958, J1604-4228 \\
          & 2017/11/21--13:33 & C43-8   & \phn92\,m -- \phn8.5\,km & 43 & 52--62 & 0.35--0.60
          & J1427-4206, J1427-4206, J1610-3958, J1604-4228 \\
Sz 114    & 2017/05/14--04:11 & C40-5   & \phn15\,m -- \phn1.1\,km & 43 & 72--73 & 1.00--1.15 
          & J1517-2422, J1427-4206, J1610-3958, J1540-3906 \\
          & 2017/05/17--02:12 & C40-5   & \phn15\,m -- \phn1.1\,km & 49 & 59--67 & 0.90--1.05
          & J1517-2422, J1517-2422, J1610-3958, J1540-3906 \\
          & 2017/09/25--23:57 & C40-8/9 & \phn41\,m -- 14.9\,km    & 42 & 29--39 & 0.80--1.10
          & J1924-2914, J1733-1304, J1610-3958, J1604-4228 \\
Sz 129    & 2017/05/14--04:11 & C40-5   & \phn15\,m -- \phn1.1\,km & 43 & 71--74 & 1.00--1.15 
          & J1517-2422, J1427-4206, J1610-3958, J1540-3906 \\
          & 2017/05/17--02:12 & C40-5   & \phn15\,m -- \phn1.1\,km & 49 & 53--61 & 0.90--1.05
          & J1517-2422, J1517-2422, J1610-3958, J1540-3906 \\
          & 2017/09/24--20:58 & C40-8/9 & \phn41\,m -- 14.9\,km    & 40 & 60--68 & 0.85--1.15
          & J1427-4206, J1427-4206, J1610-3958, J1604-4228 \\
          & 2017/11/22--11:48 & C43-8   & \phn92\,m -- \phn8.3\,km & 48 & 32--43 & 0.35--0.42
          & J1427-4206, J1427-4206, J1610-3958, J1604-4228 \\
MY Lup    & 2017/05/14--04:11 & C40-5   & \phn15\,m -- \phn1.1\,km & 43 & 69--71 & 1.00--1.15 
          & J1517-2422, J1427-4206, J1610-3958, J1540-3906 \\
          & 2017/05/17--02:12 & C40-5   & \phn15\,m -- \phn1.1\,km & 49 & 54--61 & 0.90--1.05
          & J1517-2422, J1517-2422, J1610-3958, J1540-3906 \\
          & 2017/09/24--22:14 & C40-8/9 & \phn41\,m -- 14.9\,km    & 40 & 46--57 & 0.90--1.25
          & J1427-4206, J1617-5848, J1610-3958, J1604-4228 \\
          & 2017/11/25--14:35 & C43-8   & \phn92\,m -- \phn8.5\,km & 44 & 63--68 & 0.60--0.80
          & J1617-5848, J1617-5848, J1610-3958, J1604-4228 \\
HD 142666 & 2017/09/25--21:31 & C40-8/9 & \phn41\,m -- 14.9\,km    & 42 & 56--69 & 0.80--1.00
          & J1517-2422, J1517-2422, J1553-2422, J1609-2205 \\
          & 2017/11/09--14:47 & C43-8   & 139\,m -- 13.9\,km       & 44 & 59--72 & 0.55--0.80
          & J1427-4206, J1427-4206, J1553-2422, J1609-2205 \\
HD 143006 & 2017/05/14--06:00 & C40-5   & \phn15\,m -- \phn1.1\,km & 43 & 66--79 & 0.90--1.05
          & J1517-2422, J1517-2422, J1625-2527, J1615-2430 \\
          & 2017/05/17--03:52 & C40-5   & \phn15\,m -- \phn1.1\,km & 45 & 73--80 & 0.80--0.95
          & J1517-2422, J1733-1304, J1625-2527, J1615-2430 \\
          & 2017/05/19--02:09 & C40-5   & \phn17\,m -- \phn1.1\,km & 40 & 50--66 & 0.55--0.80
          & J1517-2422, J1517-2422, J1625-2527, J1609-2205 \\
          & 2017/09/26--22:36 & C40-8/9 & \phn41\,m -- 14.9\,km    & 41 & 39--52 & 1.50--1.90
          & J1517-2422, J1733-1304, J1553-2422, J1609-2205 \\
          & 2017/11/26--13:59 & C43-8   & \phn92\,m -- \phn8.5\,km & 45 & 62--72 & 0.75--0.85 
          & J1427-4206, J1427-4206, J1553-2422, J1609-2205 \\
AS 205    & 2017/05/14--06:00 & C40-5   & \phn15\,m -- \phn1.1\,km & 43 & 66--80 & 0.90--1.05
          & J1517-2422, J1517-2422, J1625-2527, J1615-2430 \\
          & 2017/05/17--03:52 & C40-5   & \phn15\,m -- \phn1.1\,km & 45 & 73--80 & 0.80--0.95
          & J1517-2422, J1733-1304, J1625-2527, J1615-2430 \\
          & 2017/05/19--02:09 & C40-5   & \phn17\,m -- \phn1.1\,km & 40 & 50--66 & 0.55--0.80
          & J1517-2422, J1517-2422, J1625-2527, J1609-2205 \\
          & 2017/09/29--22:16 & C40-8/9 & \phn41\,m -- 15.0\,km    & 43 & 44--56 & 1.20--1.50
          & J1517-2422, J1733-1304, J1551-1755, J1532-1319 \\
SR 4      & 2017/05/14--06:00 & C40-5   & \phn15\,m -- \phn1.1\,km & 43 & 66--79 & 0.90--1.05
          & J1517-2422, J1517-2422, J1625-2527, J1615-2430 \\
          & 2017/05/17--03:52 & C40-5   & \phn15\,m -- \phn1.1\,km & 45 & 73--80 & 0.80--0.95
          & J1517-2422, J1733-1304, J1625-2527, J1615-2430 \\
          & 2017/05/19--02:09 & C40-5   & \phn17\,m -- \phn1.1\,km & 40 & 50--66 & 0.55--0.80
          & J1517-2422, J1517-2422, J1625-2527, J1609-2205 \\
          & 2017/09/06--23:08 & C40-8   & \phn41\,m -- \phn7.6\,km & 47 & 57--70 & 0.75--0.90
          & J1517-2422, J1517-2422, J1625-2527, J1633-2557 \\
          & 2017/10/17--22:42 & C43-10  & \phn41\,m -- 16.2\,km    & 47 & 27--39 & 1.40--1.85
          & J1617-5848, J1733-1304, J1625-2527, J1633-2557 \\
Elias 20  & 2017/05/14--06:00 & C40-5   & \phn15\,m -- \phn1.1\,km & 43 & 66--79 & 0.90--1.05
          & J1517-2422, J1517-2422, J1625-2527, J1615-2430 \\
          & 2017/05/17--03:52 & C40-5   & \phn15\,m -- \phn1.1\,km & 45 & 73--80 & 0.80--0.95
          & J1517-2422, J1733-1304, J1625-2527, J1615-2430 \\
          & 2017/05/19--02:09 & C40-5   & \phn17\,m -- \phn1.1\,km & 40 & 50--66 & 0.55--0.80
          & J1517-2422, J1517-2422, J1625-2527, J1609-2205 \\
          & 2017/09/23--22:35 & C40-8/9 & \phn41\,m -- 12.1\,km    & 39 & 50--63 & 0.75--0.90
          & J1517-2422, J1517-2422, J1625-2527, J1633-2557 \\
          & 2017/10/07--23:10 & C43-10  & \phn41\,m -- 16.2\,km    & 51 & 29--42 & 0.40--0.80
          & J1617-5848, J1733-1304, J1625-2527, J1633-2557 \\
DoAr 25   & 2017/05/14--06:00 & C40-5   & \phn15\,m -- \phn1.1\,km & 43 & 67--80 & 0.90--1.05
          & J1517-2422, J1517-2422, J1625-2527, J1615-2430 \\
          & 2017/05/17--03:52 & C40-5   & \phn15\,m -- \phn1.1\,km & 45 & 73--80 & 0.80--0.95
          & J1517-2422, J1733-1304, J1625-2527, J1615-2430 \\
          & 2017/05/19--02:09 & C40-5   & \phn17\,m -- \phn1.1\,km & 40 & 50--66 & 0.55--0.80
          & J1517-2422, J1517-2422, J1625-2527, J1609-2205 \\
          & 2017/09/22--23:11 & C40-8/9 & \phn41\,m -- 12.1\,km    & 40 & 42--54 & 0.65--1.10
          & J1517-2422, J1733-1304, J1625-2527, J1633-2557 \\
Elias 24  & 2017/09/25--22:42 & C40-8/9 & \phn41\,m -- 14.9\,km    & 42 & 46--58 & 0.78--1.05
          & J1517-2422, J1733-1304, J1625-2527, J1633-2557 \\
          & 2017/10/04--23:04 & C43-10  & \phn41\,m -- 15.0\,km    & 45 & 34--46 & 0.85--1.10
          & J1617-5848, J1733-1304, J1625-2527, J1633-2557 \\
Elias 27  & 2017/09/07--22:42 & C40-8   & \phn41\,m -- \phn8.8\,km & 45 & 63--76 & 1.10--1.35
          & J1517-2422, J1517-2422, J1625-2527, J1633-2557 \\
          & 2017/10/03--21:56 & C43-10  & \phn41\,m -- 15.0\,km    & 45 & 50--63 & 0.95--1.15
          & J1517-2422, J1517-2422, J1625-2527, J1633-2557 \\
DoAr 33   & 2017/05/14--06:00 & C40-5   & \phn15\,m -- \phn1.1\,km & 43 & 66--79 & 0.90--1.05
          & J1517-2422, J1517-2422, J1625-2527, J1615-2430 \\
          & 2017/05/17--03:52 & C40-5   & \phn15\,m -- \phn1.1\,km & 45 & 73--80 & 0.80--0.95
          & J1517-2422, J1733-1304, J1625-2527, J1615-2430 \\
          & 2017/05/19--02:09 & C40-5   & \phn17\,m -- \phn1.1\,km & 40 & 50--66 & 0.55--0.80
          & J1517-2422, J1517-2422, J1625-2527, J1609-2205 \\
          & 2017/09/17--23:06 & C40-8/9 & \phn41\,m -- 12.1\,km    & 46 & 47--59 & 0.80--1.20
          & J1517-2422, J1733-1304, J1625-2527, J1633-2557 \\
          & 2017/10/10--22:26 & C43-10  & \phn41\,m -- 16.2\,km    & 47 & 36--48 & 0.55--0.75
          & J1517-2422, J1733-1304, J1625-2527, J1633-2557 \\
WSB 52    & 2017/05/14--06:00 & C40-5   & \phn15\,m -- \phn1.1\,km & 43 & 66--79 & 0.90--1.05
          & J1517-2422, J1517-2422, J1625-2527, J1615-2430 \\
          & 2017/05/17--03:52 & C40-5   & \phn15\,m -- \phn1.1\,km & 45 & 73--80 & 0.80--0.95
          & J1517-2422, J1733-1304, J1625-2527, J1615-2430 \\
          & 2017/05/19--02:09 & C40-5   & \phn17\,m -- \phn1.1\,km & 40 & 50--66 & 0.55--0.80
          & J1517-2422, J1517-2422, J1625-2527, J1609-2205 \\
          & 2017/09/10--23:56 & C40-8   & \phn41\,m -- \phn7.6\,km & 43 & 43--55 & 0.50--1.00
          & J1517-2422, J1733-1304, J1625-2527, J1633-2557 \\
          & 2017/10/06--22:51 & C43-10  & \phn41\,m -- 16.2\,km    & 49 & 35--47 & 0.60--0.80
          & J1616-5848, J1733-1304, J1625-2527, J1633-2557 \\
WaOph 6   & 2017/05/09--04:28 & C40-5   & \phn15\,m -- \phn1.1\,km & 45 & 67--73 & 1.00--1.15 
          & J1517-2422, J1733-1304, J1634-2058, J1653-1551 \\
          & 2017/09/09--00:59 & C40-8   & \phn41\,m -- \phn7.6\,km & 42 & 32--45 & 0.80--1.30
          & J1751+0939, J1733-1304, J1653-1551, J1658-0739 \\
          & 2017/09/20--00:36 & C40-8/9 & \phn41\,m -- 12.1\,km    & 44 & 27--41 & 0.70--1.00
          & J1751+0939, J1733-1304, J1653-1551, J1658-0739 \\
AS 209    & 2017/05/09--04:28 & C40-5   & \phn15\,m -- \phn1.1\,km & 45 & 67--74 & 1.00--1.15 
          & J1517-2422, J1733-1304, J1634-2058, J1653-1551 \\
          & 2017/09/07--00:24 & C40-8   & \phn41\,m -- \phn7.6\,km & 46 & 42--54 & 0.60--0.85
          & J1517-2422, J1733-1304, J1653-1551, J1658-0739 \\
          & 2017/09/20--23:18 & C40-8/9 & \phn41\,m -- 12.1\,km    & 44 & 46--58 & 0.80--1.50
          & J1517-2422, J1733-1304, J1653-1551, J1658-0739 \\
HD 163296 & 2017/09/08--22:17 & C40-8   & \phn41\,m -- \phn5.8\,km & 40 & 79--87 & 1.10--1.40
          & J1924-2914, J1924-2914, J1751-1950, J1743-1658 \\
          & 2017/09/08--23:12 & C40-8   & \phn41\,m -- \phn5.8\,km & 40 & 74--87 & 1.10--1.40
          & J1924-2914, J1733-1304, J1751-1950, J1743-1658 \\
\enddata
\tablecomments{Basic information from the individual execution blocks conducted as part of ALMA Program 2016.1.00484.L.  Col.~(1) Target name.  Col.~(2) UTC date and time for the start of the execution block.  Col.~(3) ALMA configuration.  Col.~(4) Minimum and maximum baseline lengths.  Col.~(5) Number of antennas available.  Col.~(6) Target elevation range.  Col.~(7) Range of precipitable water vapor levels.  Col.~(8) From left to right, the quasars observed for calibrating the bandpass, amplitude scale, phase variations, and checking the phase transfer.  Additional archival observations used in our analysis are compiled in Table~\ref{table:obs_archive}.}
\end{deluxetable*}

\begin{deluxetable*}{lclcclcc}
\tabletypesize{\scriptsize}
\tablenum{3}
\tablecaption{Archival ALMA Datasets Used by DSHARP \label{table:obs_archive_full}}
\tablehead{
\colhead{Name} & 
\colhead{UTC Date} & 
\colhead{Config.} & 
\colhead{Baselines} & 
\colhead{$N_{\rm ant}$} & 
\colhead{Calibrators} & 
\colhead{Program} & 
\colhead{Refs.}
}
\colnumbers
\startdata
IM Lup    & 2014/07/06--22:18 & C34-4   & 20 -- \phn650 m & 31 & J1427-4206, Titan, J1534-3526, J1626-2951 & 2013.1.00226.S & 1 \\
          & 2014/07/17--01:38 & C34-4   & 20 -- \phn650 m & 32 & J1427-4206, Titan, J1534-3526, \nodata & 2013.1.00226.S & 1 \\
          & 2015/01/29--09:48 & C34-2/1 & 15 -- \phn349 m & 40 & J1517-2422, Titan, J1610-3958, \nodata & 2013.1.00694.S & 2 \\
          & 2015/05/13--08:30 & C34-3/4 & 21 -- \phn558 m & 36 & J1517-2422, Titan, J1610-3958, \nodata & 2013.1.00694.S & 2 \\
          & 2015/06/09--23:42 & C34-5   & 21 -- \phn784 m & 37 & J1517-2422, Titan, J1610-3958, J1614-3543              & 2013.1.00798.S & 3 \\
HD 142666 & 2015/07/21--22:27 & C34-7/6 & 15 -- 1600 m    & 44 & J1517-2422, Titan,                  J1627-2426, J1625-2527              & 2013.1.00498.S & \nodata \\ 
HD 143006 & 2016/06/14--03:35 & C40-4   & 15 -- \phn642 m & 37 & J1517-2422, J1517-2422, J1625-2527, \nodata & 2015.1.00964.S & \nodata \\
          & 2016/07/02--04:17 & C40-4   & 15 -- \phn704 m & 42 & J1517-2422, J1517-2422, J1625-2527, \nodata & 2015.1.00964.S & \nodata \\
AS 205    & 2012/03/27--10:08 & \nodata & 43 -- \phn402 m & 15 & J1924-2914, Titan, J1625-2527, \nodata & 2011.0.00531.S & 4 \\
          & 2012/05/04--05:11 & \nodata & 21 -- \phn402 m & 15 & 3C 279, Titan, J1625-2527, \nodata & 2011.0.00531.S & 4 \\
Elias 24  & 2015/07/21--22:27 & C34-7/6 & 15 -- 1600 m    & 44 & J1517-2422, Titan,                  J1627-2426, J1625-2527              & 2013.1.00498.S & 5 \\ 
Elias 27  & 2015/07/21--22:27 & C34-7/6 & 15 -- 1600 m    & 44 & J1517-2422, Titan,                  J1627-2426, J1625-2527              & 2013.1.00498.S & 6 \\ 
AS 209    & 2014/07/02--03:54 & C34-4   & 20 -- \phn650 m & 34 & J1733-1304, Titan, J1733-1304, \nodata & 2013.1.00226.S & 7 \\
          & 2014/07/17--02:48 & C34-4   & 20 -- \phn650 m & 32 & J1733-1304, Titan, J1733-1304, \nodata & 2013.1.00226.S & 7 \\
          & 2016/09/22--23:15 & C40-6   & 15 -- 3144 m    & 38 & J1517-2422, J1733-1304, J1733-1304, \nodata & 2015.1.00486.S & 8 \\
          & 2016/09/26--13:02 & C40-6   & 15 -- 3144 m    & 41 & J1517-2422, J1733-1304, J1733-1304, \nodata & 2015.1.00486.S & 8 \\
HD 163296 & 2014/06/04--07:10 & C34-4 & 21 -- \phn558 m & 33 & J1733-1304, J1733-1304, J1733-1304, \nodata & 2013.1.00366.S & 9 \\
          & 2014/06/14--06:20 & C34-4 & 21 -- \phn558 m & 35 & J1733-1304, J1733-1304, J1733-1304, \nodata & 2013.1.00366.S & 9 \\
          & 2014/06/16--07:09 & C34-4 & 21 -- \phn558 m & 35 & J1733-1304, J1733-1304, J1733-1304, \nodata & 2013.1.00366.S & 9 \\
          & 2014/06/17--07:25 & C34-4 & 21 -- \phn558 m & 30 & J1733-1304, J1733-1304, J1733-1304, \nodata & 2013.1.00366.S & 9 \\
          & 2014/06/29--05:44 & C34-4 & 21 -- \phn558 m & 32 & J1733-1304, J1733-1304, J1733-1304, \nodata & 2013.1.00366.S & 9 \\
          & 2015/08/05--04:15 & C34-7/6 & 42 -- 1574 m    & 37 & J1733-1304, Ceres, J1733-1304, J1812-2836 & 2013.1.00601.S & 10 \\
          & 2015/08/08--03:31 & C34-7/6 & 42 -- 1574 m    & 43 & J1733-1304, Ceres, J1733-1304, J1812-2836 & 2013.1.00601.S & 10 \\
          & 2015/08/09--00:54 & C34-7/6 & 42 -- 1574 m    & 41 & J1733-1304, Titan, J1733-1304, J1812-2836 & 2013.1.00601.S & 10 \\
\enddata
\tablecomments{Col.~(1) Target name.  Col.~(2) UTC date and time at the start of the observations.  Col.~(3) ALMA configuration.  Col.~(4) Range of baseline lengths.  Col.~(5) Number of antennas available.  Col.~(6) From left to right, the quasars observed for calibrating the bandpass, amplitude scale, phase variations, and checking the phase transfer.  \dt{An entry of `$\ldots$' indicates no calibrator was observed for checking the phase transfer.}  Col.~(7) ALMA program ID.  Col.~(8) Original references for these datasets.  Table~\ref{table:obs_archive} is published in its entirety in the electronic edition of the journal.  A portion is shown here for guidance regarding its form and content.}
\tablerefs{1 = \citet{oberg15}, 2 = \citet{cleeves17}, 3 = \citet{pinte18}, 4 = \citet{salyk14}, 5 = \citet{dipierro18}, 6 = \citet{perez16}, 7 = \citet{huang16}, 8 = \citet{fedele18}, 9 = \citet{flaherty15}, 10 = \citet{isella16}.}
\end{deluxetable*}

\end{document}